# Artificial Intelligence Studies in Cartography: A Review and Synthesis of Methods, Applications, and Ethics

Yuhao Kang[a,b], Song Gao[a*], and Robert E. Roth[a]

[a] Department of Geography, University of Wisconsin-Madison, WI, United States

[b] GISense Lab, Department of Geography, University of South Carolina, SC, United States

[*] Corresponding author: song.gao@wisc.edu

**Abstract**

The past decade has witnessed the rapid development of geospatial artificial intelligence (GeoAI) primarily due to the ground-breaking achievements in deep learning and machine learning. A growing number of scholars from cartography have demonstrated successfully that GeoAI can accelerate previously complex cartographic design tasks and even enable cartographic creativity in new ways. Despite the promise of GeoAI, researchers and practitioners have growing concerns about the ethical issues of GeoAI for cartography. In this paper, we conducted a systematic content analysis and narrative synthesis of research studies integrating GeoAI and cartography to summarize current research and development trends regarding the usage of GeoAI for cartographic design. Based on this review and synthesis, we first identify dimensions of GeoAI methods for cartography such as data sources, data formats, map evaluations, and six contemporary GeoAI models, each of which serves a variety of cartographic tasks. These models include decision trees, knowledge graph and semantic web technologies, deep convolutional neural networks, generative adversarial networks, graph neural networks, and reinforcement learning. Further, we summarize seven cartographic design applications where GeoAI have been effectively employed: generalization, symbolization, typography, map reading, map interpretation, map analysis, and map production. We also raise five potential ethical challenges that need to be addressed in the integration of GeoAI for cartography: commodification, responsibility, privacy, bias, and (together) transparency, explainability, and provenance. We conclude by identifying four potential research directions for future cartographic research with GeoAI: GeoAI-enabled active cartographic symbolism, human-in-the-loop GeoAI for cartography, GeoAI-based mapping-as-a-service, and generative GeoAI for cartography.

## 1. Introduction

In recent years, there has been an explosion in the advancement of artificial intelligence (AI) technologies, which have been employed to address a wide range of practical problems (Ertel, 2018; LeCun et al., 2015). Many scholars in GIScience are now utilizing AI to solve complex geographic problems, ranging from spatial knowledge discovery and reasoning, geographic phenomena representation and simulation, and human-environment relationship modeling (M. Chen et al., 2023; Gao, 2021; Janowicz et al., 2020; W. Li, 2020). This emerging research thrust, termed *Geospatial Artificial Intelligence* (GeoAI), is significantly enriching GIScience research (Gao, 2021; Janowicz et al., 2020). The key difference that sets GeoAI apart from generic AI lies in GeoAI's requirements for geographic knowledge and domain-specific insights. Researchers have further incorporated geographical theories or principles to guide AI models and have developed spatially-explicit GeoAI models (Kang et al., 2022; W. Li et al., 2021). Amid this backdrop, a growing number of cartography scholars have demonstrated successfully that GeoAI can accelerate previously complex cartographic design tasks and even enable cartographic creativity in new ways (Feng et al., 2019; Kang et al., 2019; Touya et al., 2019; Usery et al., 2021). Accordingly, the integration of GeoAI for cartography is now being recognized as a promising research frontier at the intersection of cartography and GIScience.

Integrating (geospatial) AI in cartography is not completely new. Its roots date at least to the early 1980s when the use of AI was employed for vectorization (Loodts, 1981). Researchers in the 1980s into the 1990s built a series of expert systems—the most popular AI approach at that time—for map design and map generalization (Forrest, 1991; V. B. Robinson et al., 1986), point-feature label placement (Christensen & Shieber, 1995), cartographic knowledge representation (Su, 1996), and map content reference (Steinhauer et al., 2001). Two other mainstream AI models emerged in the 1990s that have been leveraged widely for map design evaluation and generalization including: decision tree-based machine learning approaches (Plazanet, 1998; Steinhauer et al., 2001; Taillandier & Gaffuri, 2012; Whigham et al., 1992) and artificial neural networks (ANNs) (Allouche & Moulin, 2005; García Balboa & Ariza López, 2008; Lagrange et al., 2000; Muller, 1992; Werschlein & Weibel, 1994). Other AI approaches developed for cartographic generalization include genetic algorithms (Armstrong, 1991; Weibel et al., 1995), case-based reasoning (Keller, 1994), multi-agent system paradigm (Ruas & Duchêne, 2007), and the ant colony algorithm (C. Zheng et al., 2011). The latest boom of GeoAI for cartography is primarily motivated by the advancement of deep learning and machine learning approaches in computer science. For instance, the deep convolutional neural network (DCNN)-based approach has achieved significant improvements in image object detection tasks (Krizhevsky et al., 2017). In addition, AlphaGo, an intelligent system that uses advanced deep learning and reinforcement learning (Silver et al., 2016), beat the human world champion of the Go game, one of the most complex board games.

Deep learning and machine learning have two marked advantages for cartography. First, GeoAI techniques including deep learning and machine learning algorithms can achieve better performance in solving several complex cartographic tasks compared to classic statistical and computational approaches. For instance, GeoAI-based approaches performed better in the geographic object (e.g., buildings, road networks, map contents) identification in maps (e.g., Jiao et al., 2022b; Touya et al., 2020; Uhl et al., 2020), a topic often treated as part of data assembly and generalization within cartography. Second, GeoAI can assist cartographers in new cartographic processes that existing GIS tools are unable to tackle, enabling cartographers to enhance their creativity during design. For example, cartography is often considered both art and science, with the artistic dimension encapsulating the cartographers' creativity, ingenuity, positionality, and experience (Kraak et al., 2020). Recent breakthroughs in AI have achieved great success in modeling aspects of visual artwork including style and aesthetics (e.g., Demir et al., 2021; Jing et al., 2020; Santos et al., 2021), which then can be transferred as inspiration into map designs (Kang et al., 2019) or serve as the basis of map generalization (Feng et al., 2019; Touya et al., 2019; Usery et al., 2021).

Despite the current success of GeoAI for cartography, researchers and practitioners have growing concerns about the ethical issues of GeoAI for cartography (Griffin, 2020; B. Zhao et al., 2021a). Here, we use *ethics* to describe the rules of conduct regarding acceptable and unacceptable behavior within a particular social system (Siau & Wang, 2020), which in this case would help to diagnose and prevent misuse of GeoAI for cartography. Ethical concerns about GeoAI for cartography can be drawn from two primary arenas: logical extensions to the existing Codes of Ethics governing cartography and GIScience (British Cartographic Society, 2020; Field, 2022; GIS Certification Institute, 2022; Nelson et al., 2022), and the ethical concerns about GeoAI raised from researchers outside of cartography and GIScience (Bostrom & Yudkowsky, 2018; Jobin et al., 2019). Despite these critiques, actionable and time-tested ethical guidelines for GeoAI are limited in the literature.

Here, we provide a systematic content analysis and narrative synthesis of research studies integrating GeoAI and cartography to summarize current research and development trends regarding the usage of GeoAI for cartographic design. From this review and synthesis, we first discuss dimensions of GeoAI methods for cartography, including data sources, data formats, map evaluations, and, notably, six contemporary GeoAI models that have been employed for specific cartographic tasks: decision trees, knowledge graph and semantic web technologies, deep convolutional neural networks, generative adversarial networks, graph neural networks, and reinforcement learning. We then summarize seven cartographic design applications that have adopted GeoAI to-date: generalization, symbolization, typography, map reading, map interpretation, map analysis, and map production. We also raise five of potentially numerous ethical challenges that require consideration when integrating GeoAI and cartography: commodification, responsibility, privacy, bias, and (together) transparency, explainability, and provenance. This paper concludes with the identification of four potential research topics for future cartographic research with GeoAI: GeoAI-enabled active cartographic symbolism, human-in-the-loop GeoAI for cartography, GeoAI-based mapping-as-a-service, and generative GeoAI for cartography.

## 2. Literature Selection

We performed a content analysis and narrative synthesis of research studies integrating GeoAI and cartography following the guidelines and suggestions by prior literature review studies (Snyder, 2019; Tang & Painho, 2021; Tranfield et al., 2003). We first selected the corpus of articles by defining a search query in Scopus (Figure 1). Scopus[1] is a comprehensive scientific database that covers a wide range of disciplines, including cartography, geography, and computer science. Scopus provides access to a large number of journals, including several prestigious cartography and GIScience journals (e.g., *Cartography and Geographic Information Science* and *International Journal of Geographical Information Science*). Moreover, Scopus offers advanced search features that allow users to refine the search query onto specific keywords, titles, authors, and journal names, which helped to ensure that we captured relevant articles.

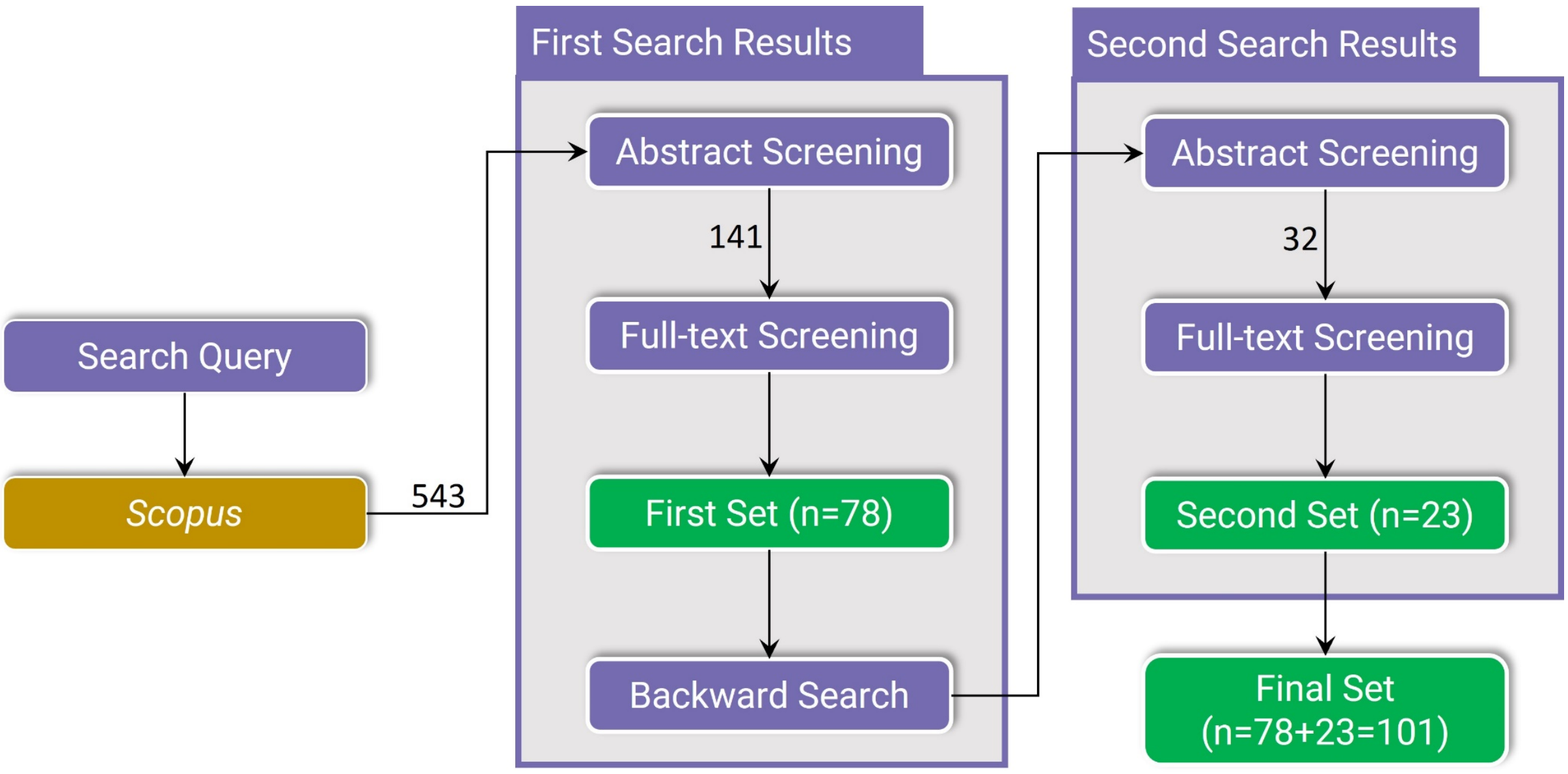

Figure 1 The article selection process for the literature review and synthesis

Our search query comprises both GeoAI methods and common cartographic design tasks:

[1] https://www.scopus.com/

*SQ = (("geoai" OR "artificial intelligence" OR "machine learning" OR "deep learning" OR "neural network" OR "knowledge graph" OR "generative adversarial" OR "reinforcement learning" OR "symbolic AI") AND ("cartography" OR "cartographic" OR "historical map" OR "map style" OR "map design" OR "map generalization" OR "map label" OR "map content" OR "participatory mapping" OR "counter mapping"))*

The initial search query resulted in 543 candidate research articles from Scopus. We performed two rounds of relevancy screening, first on abstract and then on full-text, to remove papers unrelated to the scope of our review. This two-round screening resulted in 78 candidate research articles from Scopus. After relevancy screening, we then conducted a backward search based on the manuscript reference lists from the candidate Scopus articles to include relevant studies not returned from the Scopus database. Backward search is a widely used approach in review papers (e.g., Harborth, 2017; Tang & Painho, 2021), and helped to ensure that our review provides a more comprehensive landscape of GeoAI for cartography research beyond what is found in Scopus. Notably, the backward search added several useful papers published in cartographic conference proceedings such as the *International Cartography Conference* (ICC) and *AutoCarto*. In total, we identified a final corpus of 101 papers for inclusion in our review and synthesis. These papers were all written in English and were published by December 31, 2022. A list of all considered articles in the final corpus is available as supplemental material to facilitate the transparency and reproducibility of our study.

Notably, the focus of our review is *GeoAI for cartography* rather than *GeoAI for maps* or *GeoAI for spatial data*. We excluded studies without a focus on cartographic design principles given our objective of understanding if and how GeoAI can support the cartographic design workflow. For example, we excluded studies that primarily utilized GeoAI to address bias and data quality issues in OSM data without explicitly mapping this data. Similarly, we excluded studies that focused on GeoAI for feature extraction from remotely sensed imagery, given their peripheral relevance to cartographic design decisions. By adopting this approach, our study provides a comprehensive understanding of the potential role of GeoAI in support of various cartographic design decisions.

## 3. Literature Content Analysis

We analysed the content of the sampled literature in the aggregate using several bibliometric techniques. First, we created a network map (Figure 2) of common terms from the sampled literature using VOSviewer (Eck & Waltman, 2009), a tool used in prior content analysis studies (e.g., X. Wu et al., 2022). The network map visualizes the word occurrences in the article titles and abstracts to highlight connections and visually reveal structures among concepts. We set the minimum number of word occurrences to six to emphasize the more common relationships and avoid an overwhelming number of subtopics. The most common term *neural network* appears in the center of the network map, indicating that it is the fundamental technical component of deep learning. Other notable connection points that are relevant to GeoAI techniques include *graph convolutional network* and *generative adversarial network*.

We then identified five topic clusters using a smart local moving algorithm in VOSviewer (Van Eck & Waltman, 2014; Waltman & van Eck, 2013), color coding them within the network map. The smart local moving algorithm is an optimization function designed for community detection (i.e., clustering) in large and complex networks (see Waltman & van Eck, 2013 for technical details). We adapted the smart local moving algorithm to identify semantic clusters based on the titles, abstracts, and keywords of the included papers.

The red cluster, combined with several nodes in the blue cluster, reflects articles on *map generalization,* the most common cartographic design task supported by GeoAI from the sampled literature, such as *selection*, *aggregation*, *group pattern*. The blue cluster contains terms that are relevant to cartographic *knowledge*. The green cluster includes terms such as *historical map*, *road extraction*, and *map image*, suggests the use of GeoAI such as DCNNs to identify objects from pixel and image-based raster maps. The yellow cluster refers to using *generative adversarial networks* for *map style transfer*. The purple cluster indicates the use of *graph convolutional networks* to represent *road networks* as a *graph*.

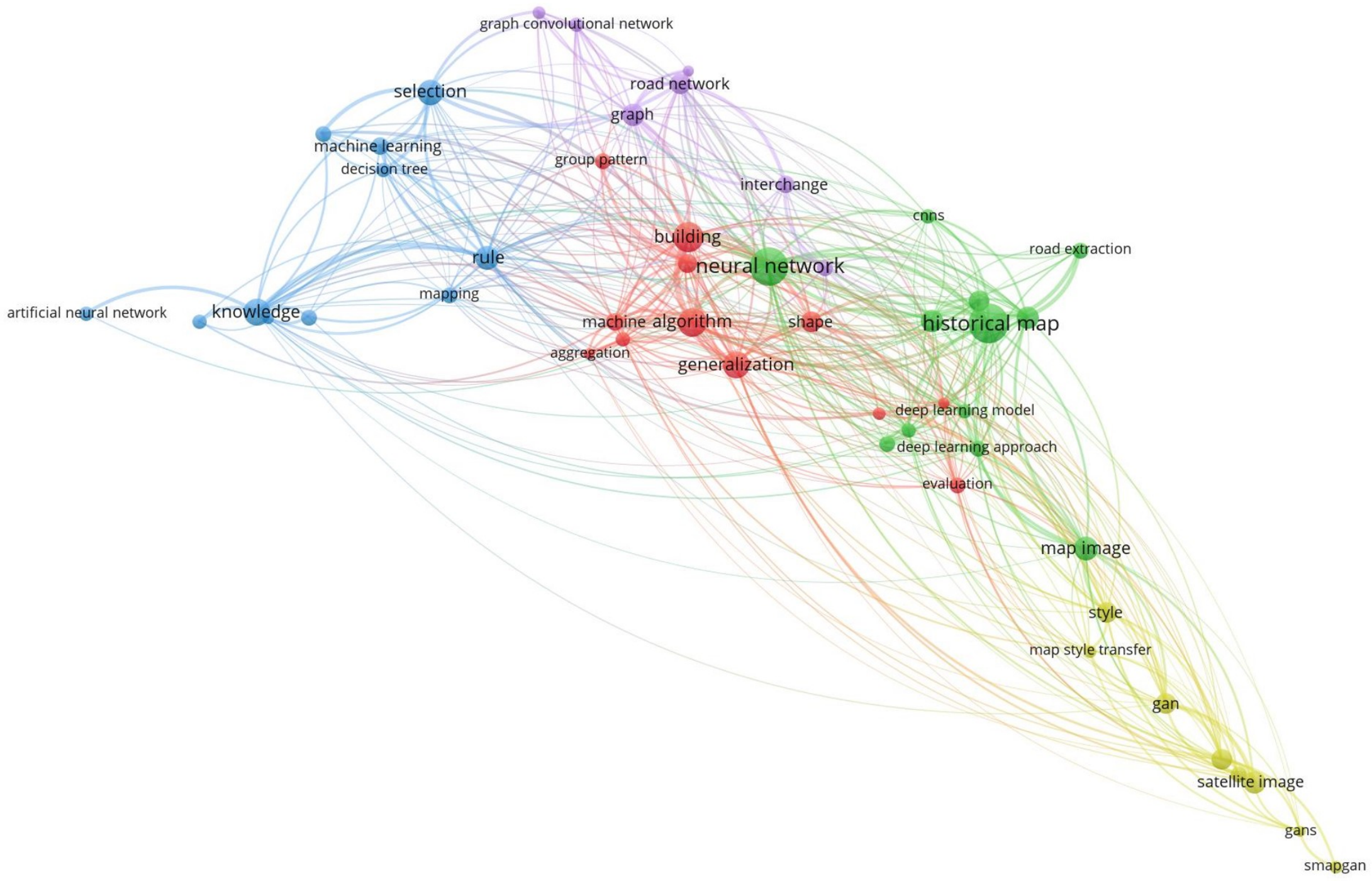

Figure 2 Network cluster map of the word occurrences found in the titles and abstracts of the papers selected for review

We drew from the network analysis to establish a working framework of research topics facing GeoAI for cartography as a way to organize and synthesize the sampled literature (Figure 3). Our framework separates GeoAI data sources, data formats, methods, and evaluations at the top and then applications to cartographic design decisions at the bottom. The GeoAI section of the framework includes six major GeoAI models employed in the sampled literature (details in Section 4). The cartography section is organized into seven categories and ten topics derived from the *Cartography and Visualization* section of *GIS&T Body of Knowledge* (2022), a comprehensive online resource that provides a foundational framework for cartography and GIScience. The *GIS&T Body of Knowledge* is written and peer-reviewed by scholars in the cartography and GIScience communities and has been widely used as a reference and textbook for educators, students, and professionals. We only discuss the cartographic design decisions observed in the sampled literature, so arguably this section of the framework could be extended to all *Cartography and Visualization* topics in the *GIS&T Body of Knowledge* as applications of GeoAI for cartography expand.

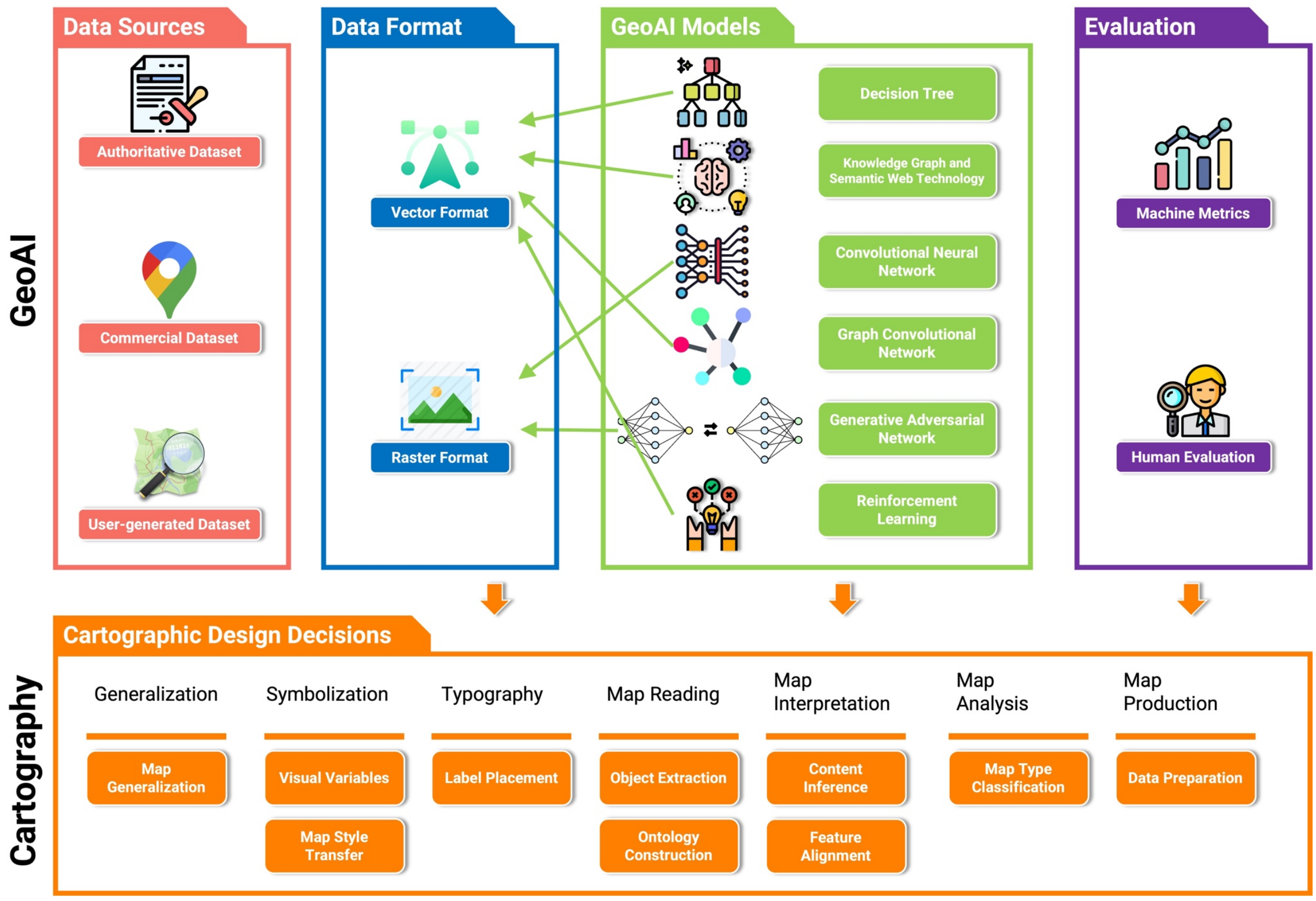

Figure 3 A conceptual framework of GeoAI in cartography

We then analyzed papers published in different years, subtopics, and venues. Figure 4 illustrates the recent rise in the integration of GeoAI for cartography. Interest in GeoAI for cartography has grown substantially, starting from only two sampled articles in 2017 to 34 papers in 2022. There is a minor decline in published work in 2021, potentially related to the impact of the COVID-19 pandemic on research productivity in 2020–2021. Overall, there appears to be a growing level of awareness and interest in GeoAI for cartography.

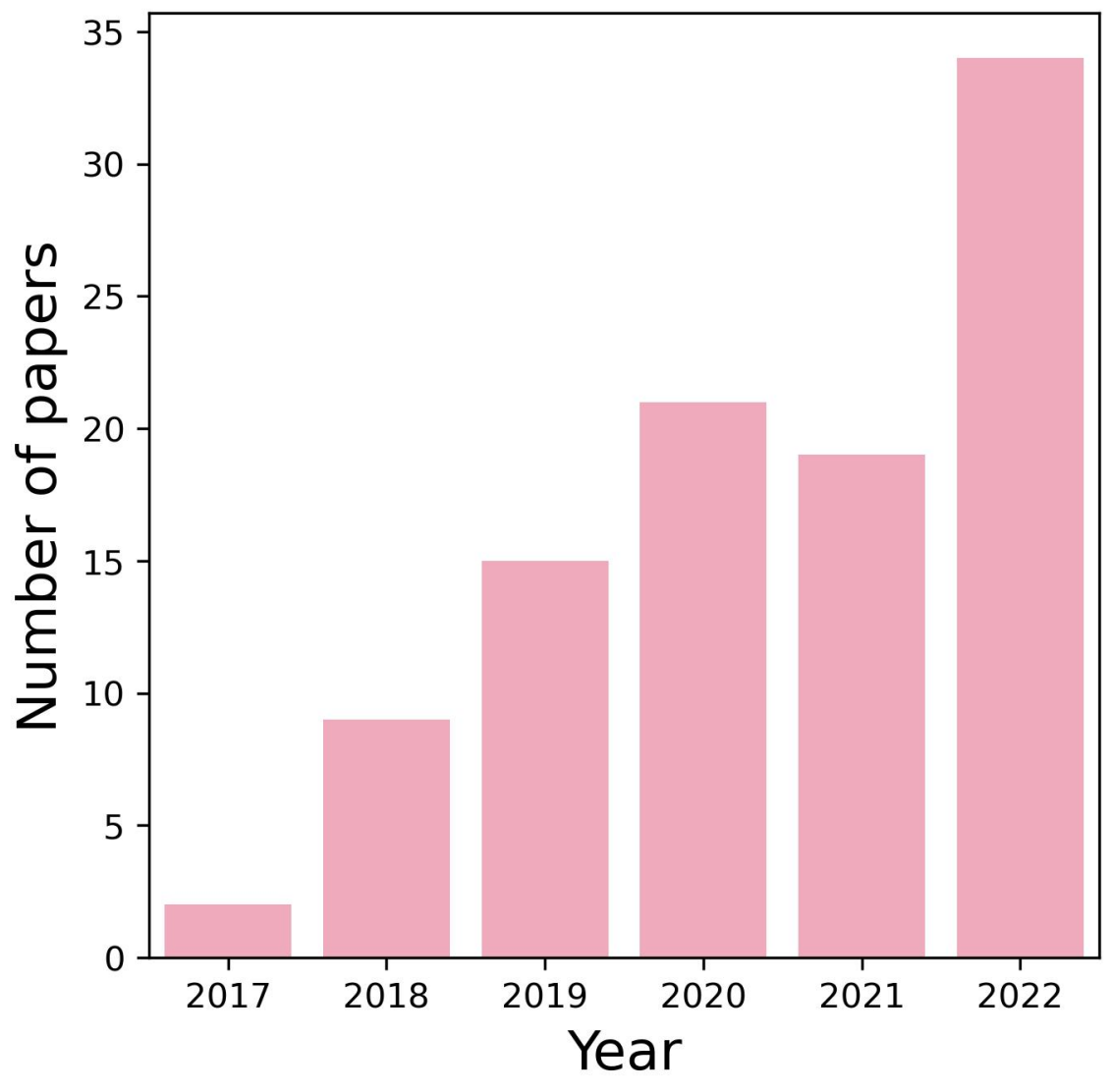

Figure 4 The total number of publications for the selected studies during the years of 2017-2022

We also analyzed the number of papers published on different cartographic design decisions derived from the *Cartography and Visualization* section of *GIS&T Body of Knowledge* (2022). As shown in Figure 5, the number of papers on each cartographic topic is unbalanced, with about a third of the sampled studies focusing on map generalization (29 papers). The second more frequent topic is map object detection (24 papers), with no other topic having over 10 papers. This summary by topic both shows where innovation currently exists regarding GeoAI for cartography, but also where future work could infill gaps in leveraging GeoAI in cartography. Notably, there are three review papers in the sample that have a narrower scope than our review and synthesis presented here. Two of these review papers focus on a specific cartographic design decisions such as geographic feature recognition (or map object detection) (Chiang et al., 2020) and road network extraction (Jiao et al., 2021). The third review paper summarizes several practical applications of GeoAI for topographic mapping in the United States Geological Survey (Usery et al., 2021).

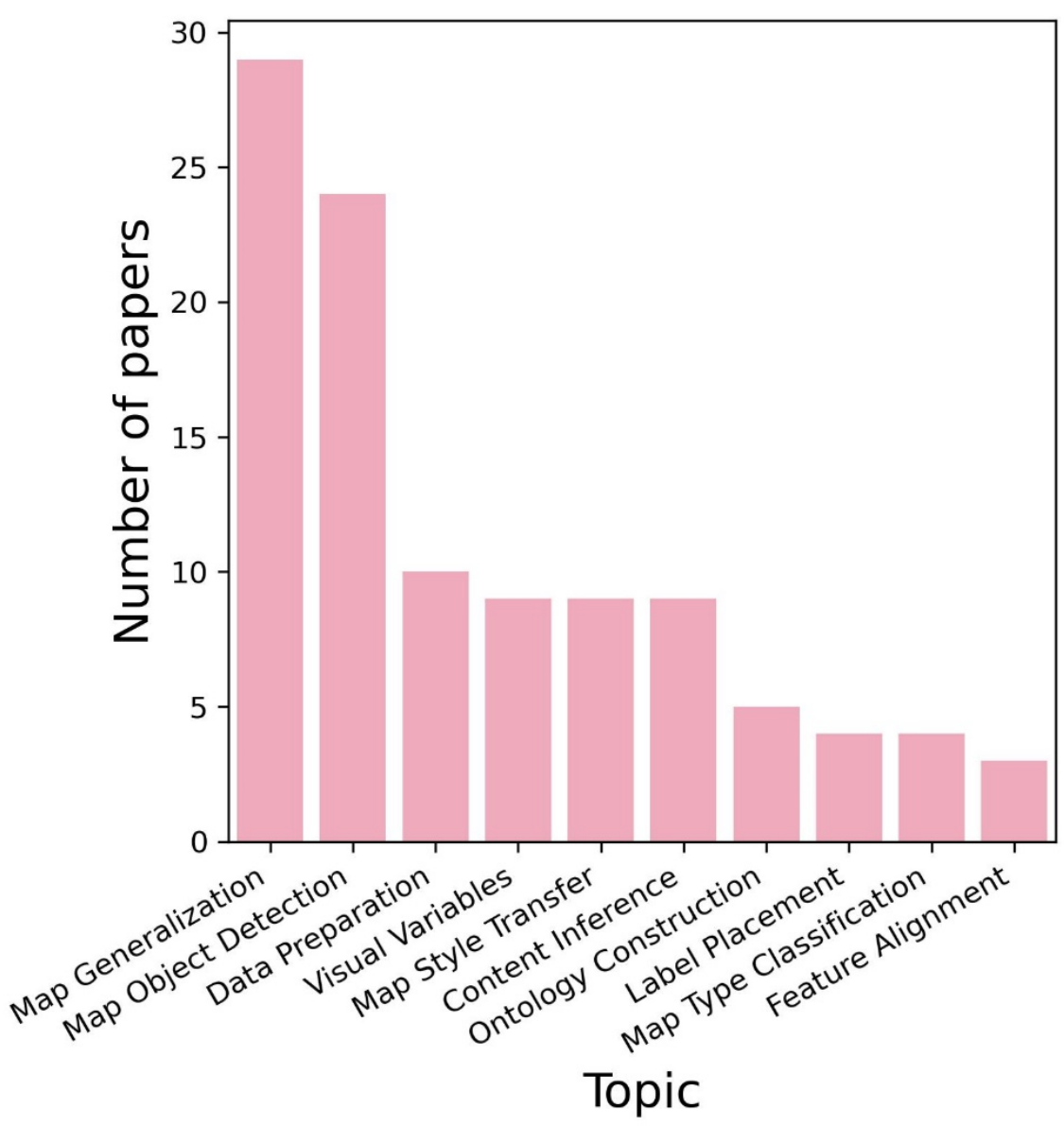

Figure 5 The number of publications of the selected studies by topic

We also analyzed the publication outlets of sampled literature. Of the final set of 101 papers, 71 (70%) were published in peer-reviewed journals and 28 (28%) in conference proceedings. Figure 6 shows the journals that published at least two papers from the set of reviewed papers. Among all journals, the *ISPRS International Journal of Geo-Information* published the most papers (ten) on GeoAI for cartography. It is closely followed by *International Journal of Geographical Information Science* (nine), and *Cartography and Geographic Information Science* (eight). Other cartography and GIScience journals including *Transactions in GIS*, *IEEE Access, International Journal of Cartography*, and *Journal of Geovisualization and Spatial Analysis* also published over three papers. In addition to these peer-reviewed journals, we expanded our research to include relevant conferences in the field. These primary conferences include but are not limited to *AutoCarto*, the *ICC (International Cartographic Conference)*, and *ACM SIGSPATIAL*. Also, seven conference papers were published in *The International Archives of the Photogrammetry, Remote Sensing and Spatial Information Sciences*. This summary gives researchers and professionals a sense of where to stay up-to-date on the latest developments on GeoAI for cartography, and also which publication outlets may be most amendable to publishing future work.

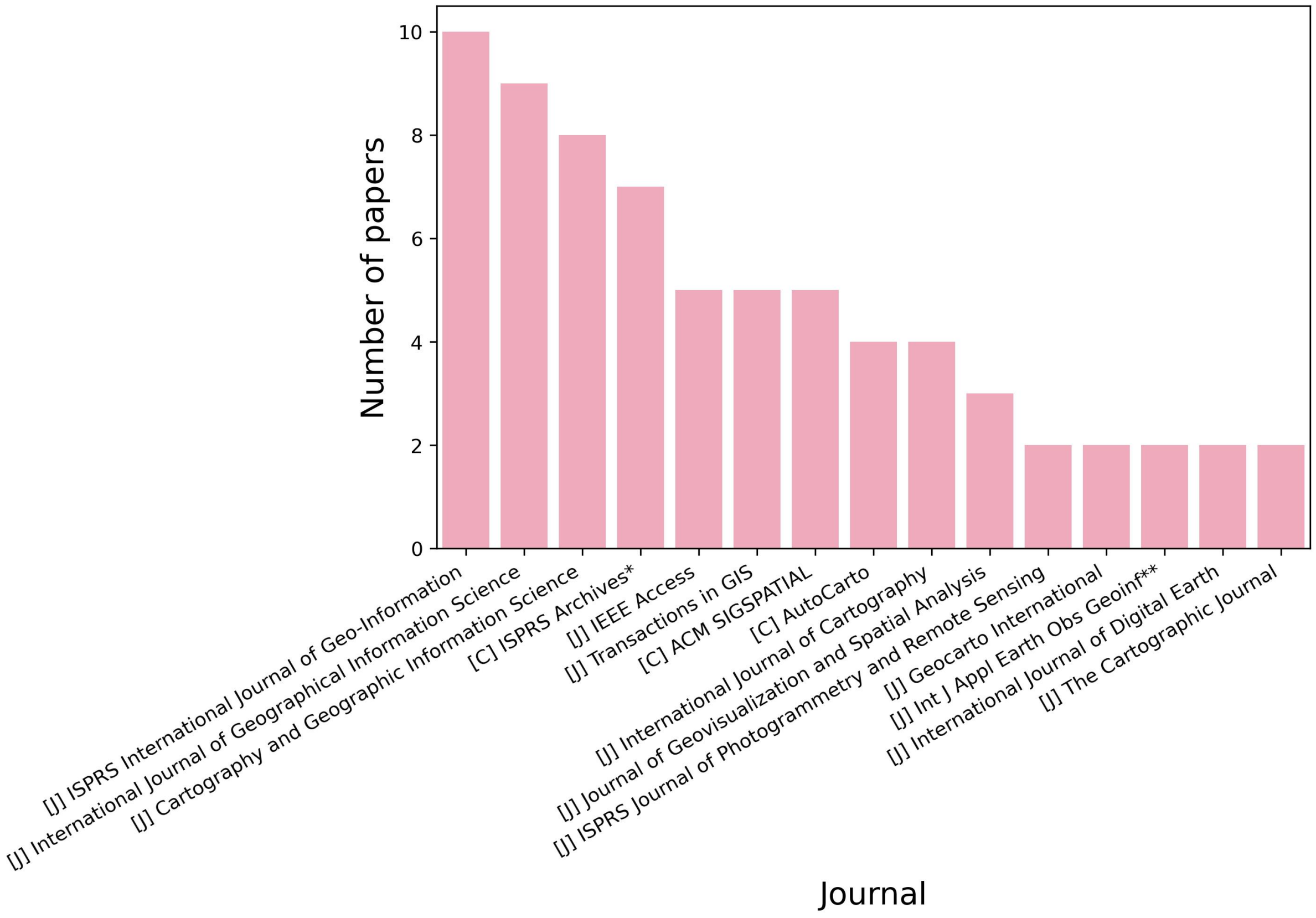

* The International Archives of the Photogrammetry, Remote Sensing and Spatial Information Sciences.

** International Journal of Applied Earth Observation and Geoinformation

Figure 6 Journals and conference proceedings that published at least two papers from the final set for review. [J] indicates journals and [C] indicates conference proceedings.

Finally, we analyzed co-authorship networks to understand the geographic distribution of researchers working on GeoAI for cartography. According to VOSviewer, 15 groups of researchers from seven countries (China, France, Germany, Poland, Sweden, Switzerland, and the United States) have published at least two papers on GeoAI for cartography. We summarize their majority affiliates, key themes, and one representative publication in Table 1. The authorship network summary is valuable for the future connecting the international community of researchers, professionals, and policymakers working on GeoAI for cartography.

Table 1 Research teams that were working on the integration of GeoAI for cartography studies.

| **Country** | **Affiliates** | **Key themes** | **Representative Publication** |
|---|---|---|---|
| **China** | Central South University | Data preparation | X. Chen et al., (2021) |
| | Information Engineering University | Map generalization | Du, Wu, Yin, et al., (2022) |
| | Wuhan University and China University of Geosciences | Map generalization | Yan et al., (2019) |
| | Wuhan University | Map type classification, data preparation | Hu et al., (2021) |
| **France** | IGN | Map generalization | Touya et al., (2019) |
| **Germany** | Leibniz University Hannover | Map generalization | Feng et al., (2019a) |
| **Poland** | University of Warsaw | Map generalization | Karsznia & Weibel, (2018) |
| **Switzerland** | ETH Zurich | Map style transfer, map object detection | Jenny et al., (2021) |
| **Sweden** | Lund University | Cartographic knowledge representation | Huang & Harrie, (2020) |
| **United States** | University of Colorado, Boulder and University of Minnesota | Map object detection | Uhl et al., (2020) |
| | University of Washington | Ethics | B. Zhao et al., (2021) |
| | University of Wisconsin-Madison | Map style transfer | Kang et al., (2019) |
| | U.S. Geological Survey | GeoAI for cartography | Usery et al., (2021) |

## 4. Literature Synthesis: Data Sources, Models, and Evaluations

### *4.1. Data Sources*

Large-scale high-quality data is the prerequisite to training a GeoAI model. In this section, we characterize three types of data sources: authoritative datasets, commercial datasets, and user-generated datasets.

*Authoritative datasets* are maintained and released by governments, such as the U.S. Census Bureau and similar mapping agencies around the world. Such datasets are considered high in data quality and commonly are employed for cartographic design. However, the geographic coverage of authoritative datasets often is restricted to specific governmental regions and scales, which limit cartographic design for cross-border projects. Furthermore, accessing authoritative datasets may be managed by different mapping agencies within a single governmental unit, limiting the harmony of authoritative datasets across agencies. Notably, several governmental mapping agencies provide historical archives that allow cartographers to analyze and extract geographic information from historical maps. Such historic archives may be relevant for training GeoAI models. However, historical archives often are limited in access and restricted to specific map types and regions. For instance, historical archives typically concentrate on specific categories like topographic maps that provide rich information on physical and human-made features but may lack thematic data (e.g., maps depicting historic climates or population distributions).

In contrast, *commercial datasets* are provided by large technology companies such as Google Maps and Bing Maps. These mapping companies offer relatively high-quality datasets, making them valuable resources for cartographic design. However, these mapping companies may charge users to access the data or have variable restrictions depending on different terms of usage for their services. There are also potential copyright issues when using commercial datasets for GeoAI, which relates to ethical discussions regarding commodification below. Researchers and professionals need to carefully evaluate the copyright and licensing policies related to these commercial datasets to ensure their ethical usage for GeoAI and cartography.

Finally, *user-generated datasets* refer to spatial data that is generated by users actively (regarding *volunteered* data) or unconsciously (regarding *contributed* data) (Harvey, 2013). OpenStreetMap (OSM) is one of the most representative open-sourced and user-generated mapping services that allows users to contribute geospatial data around the globe. In addition, researchers also have suggested using search engines and social media to collect maps for specific regions (Evans et al., 2017; Hu et al., 2021; A. C. Robinson, 2019; Schnürer et al., 2021). User-generated datasets encourage community participation and therefore can offer valuable insights into localized geographic knowledge. Furthermore, such datasets often serve as a decentralized sources of map knowledge, which could promote more inclusive and participatory research and practice in cartography. However, user-generated datasets also come with limitations when used in GeoAI for cartography. One challenge is the quality and reliability of the dataset, as contributions from multiple users may lead to inaccuracies, incomplete information, and biases. For instance, these datasets often lack standardized data structures and formats, and they also may exhibit spatial and temporal biases as they may overlook specific demographics and less digitally connected populations. Therefore, it is necessary to perform quality control measures to verify the accuracy of the user-generated dataset. A second challenge relates to privacy, particularly for data collected unconsciously, as discussed in 6.3.

Table 2 summarizes the openly-available datasets that have been used to train GeoAI models discussed in our sampled literature. GeoAI models are only as useful as the datasets used to train them, and therefore understanding differences in available datasets is important for leveraging GeoAI for cartography.

Table 2 Data sources that have been utilized to train GeoAI models for cartography

| Data source type | Dataset name | Data source | Country | Example publication | Website |
|---|---|---|---|---|---|
| Authoritative | Historical Topographic Map Collection | U.S. Geological Survey | United States | Uhl et al., (2022) | https://www.usgs.gov/programs/national-geospatial-program/historical-topographic-maps-preserving-past |
| | National Transportation Dataset | U.S. Geological Survey | United States | Uhl et al., (2022) | https://data.usgs.gov/datacatalog/data/USGS:ad3d631d-f51f-4b6a-91a3-e617d6a58b4e |
| | Geoportal | French national map agency (IGN) | French | Christophe et al., (2022) | https://www.geoportail.gouv.fr/ |
| | Lantmäteriet maps Sweden | Swedish Mapping, Cadastral and Land Registration Authority | Sweden | Ståhl & Weimann, (2022) | https://www.lantmateriet.se/sv/geodata/ |
| | TOPNL | The Netherlands' Cadastre, Land Registry and Mapping Agency | Netherland | Kang et al., (2021) | https://www.pdok.nl/introductie/-/article/basisregistratie-topografie-brt-topnl |
| | Database of General Geographic Objects (BDOO) | Head Office of Geodesy and Cartography | Poland | Lisiewicz & Karsznia, (2021) | https://www.geoportal.gov.pl/en/dane/baza-danych-ogolnogeograficznych-bdo |
| | WWII Topographic Maps | Digital Archive @ McMaster University Library | World | Ekim et al., (2021) | http://digitalarchive.mcmaster.ca/islandora/object/macrepo%3A32223 |
| | EuroGlobalMap | Europe's National Mapping, Cadastral and Land Registration Authorities | Europe | Lisiewicz & Karsznia, (2021) | https://eurogeographics.org/ |
| | Topographic Atlas of Switzerland (Siegfried map) | the Swiss national mapping agency | Switzerland | S. Wu et al., (2022) | https://www.swisstopo.admin.ch/en/geodata/maps/historical.html |
| | INSPIRE Directive | Infrastructure for Spatial Information in Europe | Europe | Z. Yang et al., (2020) | https://inspire.ec.europa.eu/Themes/Data-Specifications/2892 |
| | GB1900 | National Library of Scotland | British | Z. Li et al., (2021) | https://geo.nls.uk/maps/gb1900/ |
| Commercial | Google Maps | Google | World | Kang et al., (2019) | https://www.google.com/maps |
| User-generated | OpenStreetMap | - | World | Y. Xu et al., (2019) | https://www.openstreetmap.org/ |
| | Map History / History of Cartography | - | World | Schnürer et al., (2021) | https://www.maphistory.info/ |
| | Search engine | - | World | Evans et al., (2017) | - |
| | Social Media Platform (e.g., Pinterest, Twitter) | - | World | Schnürer et al., (2021) | - |

### *4.2. Data Formats and GeoAI Models*

Vector and raster are the two primary data formats used for storing and representing geospatial data for cartography (Peter & Weibel, 1999) and also are the two primary input data formats for GeoAI models (Janowicz et al., 2020). *Vector map data* treat each geographic element as a distinct object comprising points, polylines, or polygons. These objects contain both geometry and attributes, and can be manipulated using a variety of spatial operations. *Raster map data*, in contrast, are stored as images, with each pixel corresponding to a spatial area with the digital number associated with the pixel storing geographic information about that area.

Among all the reviewed papers, we identified six categories of GeoAI models that have been widely used in the current cartographic studies: decision trees (DT), knowledge graph and semantic web technologies (KG & SWT), deep convolutional neural networks (DCNNs), generative adversarial networks (GANs), graph neural networks (GCNs), and reinforcement learning (RL). It is important for researchers and professionals to consider the unique characteristics of each model, particularly those that make them more or less appropriate to vector vs. raster input and output map data formats during cartographic design.

A *Decision Tree* (DT) is a tree-like model in which each node denotes an attribute, each branch represents a choice, and each leaf indicates a class label. Several notable decision tree-based machine learning approaches include random forests (Breiman, 2001), Gradient boosting machines (Friedman, 2001), AdaBoost (Hastie et al., 2009), and XGBoost (T. Chen & Guestrin, 2016). Decision trees are appropriate for classifying input data based on a set of input attributes, making them well-suited for processing vector map formats wherein each geographic object is represented as an individual entity with several attributes. For instance, researchers have utilized decision trees for settlement selection (Lisiewicz & Karsznia, 2021) by determining which settlements should be included or excluded based on various variables including population, types of settlements, and functions. Researchers have leveraged decision trees for other cartographic design decisions as well, including map generalization (Lisiewicz & Karsznia, 2021) and color quality evaluation (T. Chen et al., 2021). More importantly, several advanced decision tree models explicitly have been employed for cartography, including random forests (He et al., 2018) and AdaBoost (T. Chen et al., 2021). These models, which combine multiple decision trees, tend to yield better performance in solving cartographic design tasks. One key strength of decision trees lies in their interpretability, as researchers can identify what variables contribute to cartographic design outcome. However, decision tree models may require manual parameter tuning and may exhibit overfitting issues. Also, decision tree models primarily rely on attributes to produce results while spatial characteristics are overlooked. In the future, given the rising concerns regarding transparency and interpretability of the GeoAI for cartography, decision tree models may offer valuable insights in supporting cartographic design decisions.

*Knowledge Graph* and *Semantic Web Techniques* (KG & SWT) have been utilized to summarize cartographic knowledge. Researchers have employed these techniques to build ontologies and semantic design rules, such as demonstrating the relationships among a variety of visual variables in maps and geovisualizations. It is possible to further support logistic reasoning such as question answering (Huang & Harrie, 2020; Mai et al., 2022a) and map feature linking (Shbita et al., 2020) via KG & SWT. For instance, Huang

& Harrie, (2020) first constructed an ontology structure to encompass map elements such as style, symbol, and legend. They subsequently developed a system based on this ontology, empowering users to retrieve geographic information such as spatial relationships from the system. One primary advantage of KG & SWT is their capacity for reasoning, which provides a structured and semantic-based method to extract spatial knowledge from geographic data. Using KG & SWT, cartographers could delve deeper into the insights offered by maps and visualizations. However, a potential limitation of KG & SWT refers to the requirement for predefined ontological structures. Given the subjectivity inherent in several cartographic design decisions, it is intractable to define objective structures that represent all cartographic design decision processes. In the future, KG & SWT may play an increasingly important role in evaluating GeoAI for cartography. In particular, KG & SWT, with their inherent logistics and reason-based approaches, may minimize potential risks for GeoAI to generate unethical results such as fake maps and other potential negative outputs.

*Deep Convolutional Neural Networks* (DCNNs) are used widely to analyze images, and have achieved high performance in multiple computer vision applications such as image classification, image object detection and localization, and image segmentation (Aloysius & Geetha, 2017). Given their inherent adeptness at image processing, they are particularly appropriate for raster-based map processing. DCNNs are developed based on artificial neural networks (ANNs) with multiple layers. DCNNs contain filters or kernels that act like a sliding window to learn high-level visual features from neighboring cells of the input images. Researchers have adapted DCNNs to solve cartographic applications such as map type classification (Z. Yang et al., 2020), map feature detection (Chiang et al., 2020), and map generalization (Feng et al., 2019). For instance, Feng et al., (2019) trained an end-to-end DCNN model to produce generalized maps of building footprints. Provided with datasets of original and manually generalized maps, DCNNs can learn the generalization patterns between datasets and then apply the generalization solutions to new, ungeneralized maps. The major strength of DCNNs is their ability to extract complex visual and spatial features from map images. Also, DCNNs achieve higher performance in most image processing tasks than alternative GeoAI models. Nevertheless, DCNNs are not without their limitations. In general, building DCNNs requires a large amount of labeled training dataset and extensive computational resources (e.g., high-performance computing and numerous GPUs). They also are inherently "black-box" models, making it difficult to interpret and reproduce the modeling process, potentially raising ethical and transparency considerations. Looking towards the future, more advanced DCNNs (e.g., vision transformers) could be developed to offer deeper understanding of maps with superior performance.

*Graph Convolutional Networks* (GCNs) are a generalized version of DCNNs that are appropriate for modeling irregularly structured data. Vector data can be represented as graphs that contain nodes and edges. For points and polygons, graphs can be constructed based on the spatial adjacency relationships of the geographic objects and their neighbors. Polylines such as road and river networks effectively are graphs and therefore work well with GCNs. GCNs have been used for modeling vector-format data based on their constructed graph structures. GCNs can learn high-level features and spatial relationships of geographic objects by considering the centering nodes and their neighbors. Cartographers have leveraged GCNs to learn the representations of geographic objects (X. Yan et al., 2021; M. Yang, Jiang, et al., 2022) and perform map generalizations (X. Yan et al., 2019). For instance, in the study by Yan et al., (2019), a graph was first constructed to represent the spatial adjacency relationships among buildings. Then, each building was treated as a node, encoded with multiple attributes, and input into the GCNs to determine the building group patterns. One major advantage of GCNs lies in their capacity for modeling irregular vector data, offering an improvement over DCNNs that primarily are designed for grid-like data structures. In the future, GCNs may play a key role in embedding cartographic principles and spatial patterns into GeoAI models.

*Generative Adversarial Networks* (GANs) are another popular approach in the cartography community. GANs contain two components—the generator and a discriminator—each of which works as a DCNN. Often, the generator creates synthetic data and the discriminator then is used to judge whether the outputs created by the generator exhibit similar patterns as the input data. The two components are trained together until the discriminator cannot differentiate the real input map images from the generated synthetic data, resulting in plausible generated maps. Existing cartographic studies that use GANs are primarily built based on DCNNs. Thus, GANs are also appropriate for raster map processing. Given its promise of producing plausible map designs, cartographers have utilized GANs for automating raster map processing tasks like map generalization (Feng et al., 2019) as well as the transfer of artwork styles and aesthetics to map tilesets (Christophe et al., 2022; Kang et al., 2019). For instance, Kang et al., (2019) trained a GAN model that learned map styles from target styled maps (e.g., Google Maps) and subsequently transferred to simple styled maps (e.g., unstyled OSM data). The GAN iteratively generated styled maps until the discriminator could not differentiate between the target styled maps and transferred styled maps. A key strength of GANs refers to their capability to generate new map designs that do not currently exist, thereby encouraging cartographic creativity. However, given that GANs may create unrealistic artifacts, and, in the worst case, result in “deep fake geography” (B. Zhao et al., 2021b). Also, GANs often require even more complex training and parameter tuning than DCNNs. In the future, GANs might be utilized to generate artistic features of maps, which may serve as inspiration during cartographic design.

Finally, *Reinforcement Learning* (RL) approaches generate optimal map solutions by rewarding positive actions while penalizing undesirable ones. Duan et al., (2020) aligned vector objects to the corresponding map features in rasterized historical maps using reinforcement learning. Despite the rapid development of reinforcement learning outside of the GIScience community, the Duan et al., (2020) study is the only example of reinforcement learning approaches for cartography in our sampled literature. The primary strength of RL lies in its capability to learn and optimize sequences of actions, which makes RL particularly suitable for addressing complex cartographic designs that involve a series of interdependent decisions. However, because RL requires well-defined reward functions and actions, it is challenging to assign the rewards and penalties, particularly given the multifaceted and subjective nature of cartographic design. Moreover, RL models are computationally expensive and require a large amount of training data. These constraints might restrict the broad applicability of RL within the field of cartography, although potential may still exist for complex and multi-action decision-making such as multiscale map generalization that employs a range of generalization operators (e.g., Roth et al., 2011) versus just focusing on selection or simplification.

Table 3 summarizes the relative pros and cons of the reviewed data formats and GeoAI models. Figure 7 shows the number of papers that employed each GeoAI model. DCNNs are among the most popular GeoAI models (49 papers), followed by GANs (16). Thus, the interrelated DCNNs and GANs have been explored for cartographic applications in over 60 papers since 2017, indicating that deep learning is a growing area of interest in cartography. The remaining GeoAI models are relatively underexplored, and therefore might indicate an untapped opportunity for future research on GeoAI for cartography.

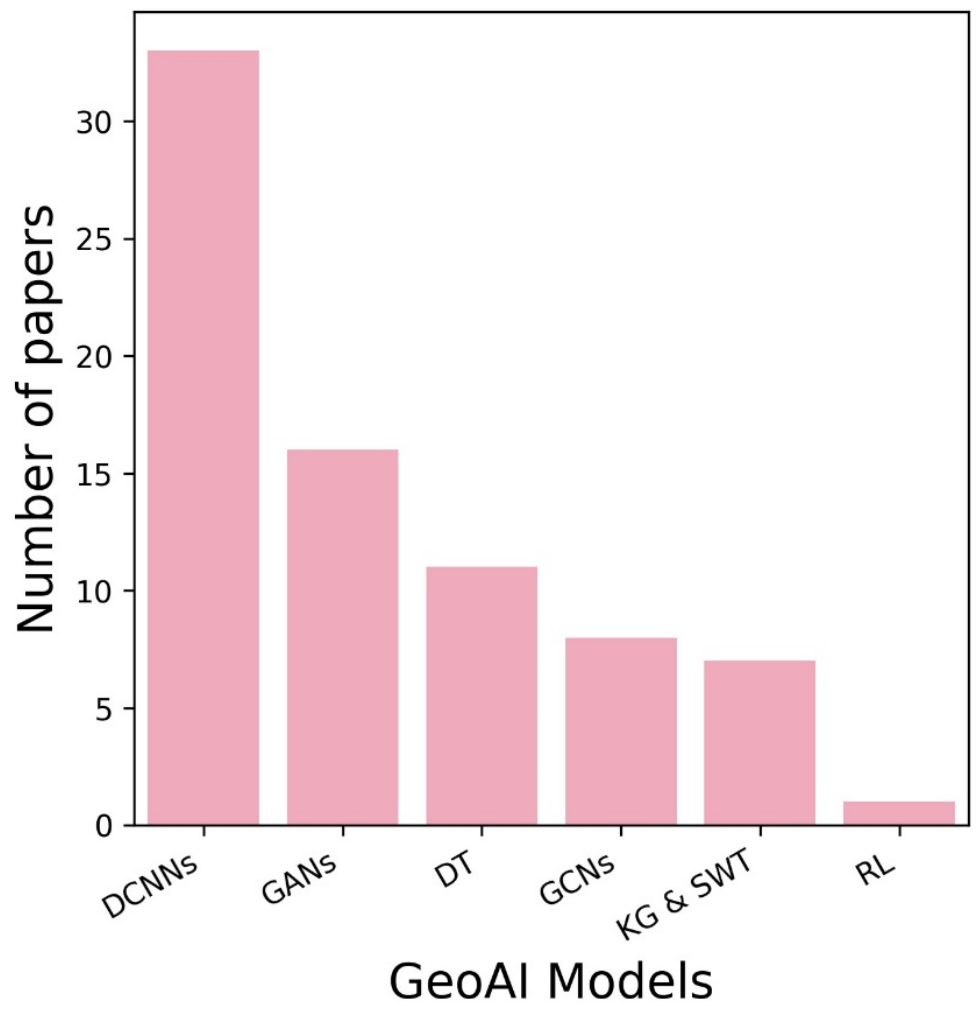

Figure 7 Number of sampled articles using each GeoAI model.

Table 3 An overview of the characteristics (including data formats, pros and cons) of six GeoAI models.

| GeoAI Models | Primary Data Formats | Pros | Cons |
|---|---|---|---|
| **Decision Tree** | Vector | Good interpretability | 1. Require manual parameter tuning<br>2. Overfitting issues<br>3. Overlook spatial characteristics |
| **KG & SWT** | Vector | Support reasoning | Predefined ontological structures |
| **DCNNs** | Raster | 1. Extract complex visual and spatial features<br>2. Higher performances | 1. Require a large amount of labeled training dataset<br>2. Extensive computational resources<br>3. "Black-box" models |
| **GCNs** | Vector | Capacity for modeling irregular vector data | 1. Require large amount of labeled training dataset<br>2. Extensive computational resources<br>3. "Black-box" models |
| **GANs** | Raster | Capability to generate new map designs | Complex training and parameter tuning |
| **RL** | Vector | Address complex cartographic designs that involve a series of interdependent decisions | 1. Require well-defined reward functions and actions<br>2. Require a large amount of labeled training dataset |

| | |
|---|---|
| | 3. Extensive computational resources |

### *4.3. Evaluation*

GeoAI also can evaluate the output map designs from a deployed model. GeoAI for evaluation serves much like a machine-based map design critique, which may present new possibilities for improving the quality and efficacy of cartographic design, particularly for complex projects with numerous map sheets or multiscale design across multiple zoom levels.

We observed in the sampled articles two primary ways to use GeoAI for cartographic evaluation. First, researchers have defined a set of machine learning metrics to evaluate the performance of different GeoAI models. Metrics may be adopted from computer science, such as Recall, Precision, and MAE (mean absolute error) (Feng et al., 2019; M. Yang, Jiang, et al., 2022), or consider new cartographic design constraints such as smoothness, coalescence reduction (Courtial et al., 2022a), position error, and size change (M. Yang, Yuan, et al., 2022). Wu et al., (2022) further modeled the uncertainties of image segmentation models in historical maps. In addition to calculating these metrics, researchers also have developed GeoAI models for evaluating the visual cartographic output (Kang et al., 2019; M. Yang, Yuan, et al., 2022). These quantitative machine-based metrics allow cartographers to evaluate the large-scale cartographic outputs produced by GeoAI models, particularly when applied across a global tileset at multiple scales. Nonetheless, these metrics have several limitations. First, given the subjectivity and complexity of many cartographic design decisions, it becomes challenging to encode them accurately using these metrics. As a result, these metrics may only provide a shallow evaluation of the output without in-depth insights into specific cartographic design decisions. Second, even though these metrics can be easily computed by machines, it might be hard for users to interpret these metrics and make meaningful cartographic design adjustments based on the metrics. Users may find it challenging to understand the significance of minor changes in these metrics, as they may not align with their prior experience and preferences for cartographic design.

Second, GeoAI enables human-in-the-loop modeling to include cartographers in visual design assessment. Despite the efficiency of machine learning metrics for cartographic evaluation, there is no objective "ground truth" dataset for most cartographic design decisions (Courtial et al., 2021). Thus, robust evaluations still require some human observation of and intervention into the output quality of maps. However, cartographers can guide GeoAI models to evaluate map qualities through visual assessments (T. Chen et al., 2021; Karsznia, Wereszczyńska, et al., 2022), effectively scaling the abilities of humans using GeoAI. For instance, Armstrong, (2019) suggested training deep learning (i.e., DCNN) approaches to evaluate dot map quality, using the model to select the candidate maps for human evaluation.. Robinson (2019) performed a content analysis of maps by extracting objects and analyzing visual contents with computer vision techniques. Similarly, Dobesova (2020) retrieved content and metadata from maps and matched maps that have similar visual styles. These assessments enabled by GeoAI are similar to map design critiques, which are commonly employed to evaluate a map's design choices and its ability to effectively deliver geographic information to the target audience. Therefore, GeoAI has the potential to complement conventional cartographic design critiques and enhance their efficiency. However, a drawback to human-in-the-loop GeoAI for cartography is that it requires expert labor. Further, different experts bring unique perspectives and preferences, which will be averaged by human-in-the-loop GeoAI evaluations and therefore may not reward the more innovative designs. However, human-in-the-loop GeoAI still reduces the necessary expert labor for large-scale mapping projects, and may lead to more satisfying results since cartographers can still override GeoAI-driven metrics in support of their cartographic design vision.

## 5. Literature Synthesis: GeoAI Applications for Cartography

In this section, we list emerging applications of GeoAI for solving cartographic design decisions. As introduced above, we follow the *Map Design Fundamentals* and *Map Use* entries in the *Cartography and Visualization* section of *GIS&T Body of Knowledge* (2022) to classify these GeoAI applications for cartography. A sampled article can belong to multiple categories because a number of papers have utilized GeoAI for multiple cartographic tasks. We start each section by providing an overview of the cartographic design decision as summarized in the *GIS&T Body of Knowledge*, followed by a summary of GeoAI applications supporting that cartographic design decision.

### *5.1. Scale & Map Generalization*

Map generalization is an important cartographic design decision that has attracted researchers' attention for decades (A. H. Robinson et al., 1995), and was the most common GeoAI application in our sampled articles. *Generalization* refers to the process of abstracting and transforming geospatial data in order to meaningfully reduce their detail (Raposo, 2017). Map generalization tasks are achieved through application of generalization operators (e.g., simplification, smoothing, and aggregation) to typically vector map data (e.g., points, polylines, and polygons) (R. E. Roth et al., 2011). AI was used for map generalization as early as the 1980s and researchers have employed deep learning for automating map generalization since 2018 (Feng et al., 2019; Sester et al., 2018). Touya et al., (2019) discuss the opportunities and challenges of GeoAI approaches for cartographic generalization.

The most common generalization operator for point-based generalization is *selection*, or choosing a subset of representative features from a given point pattern. Decision trees have been employed for point selection, primarily for human settlement selection based on population or other relevant attributes (Izabela & Karolina, 2020; Karsznia, Weibel, et al., 2022; Karsznia & Weibel, 2018; Lisiewicz & Karsznia, 2021).

In contrast, researchers have used DCNNs, GCNs, GANs, as well as decision trees for more complex polyline-based generalization solutions. These GeoAI models have been employed for the selection and simplification of road networks (Karsznia, Wereszczyńska, et al., 2022), mountain roads (Courtial et al., 2020), river networks (C. Yan et al., 2022), and coastlines (Du, Wu, Xing, et al., 2022). Yu & Chen, (2022) and Du, Wu, Yin, et al., (2022b) organized polylines as a set of points and used the points as input into neural networks to determine whether a vertex between two adjacent points should be retained or deleted. Researchers have utilized decision trees (Karsznia, Wereszczyńska, et al., 2022; Y. Xu et al., 2019) and GCNs (J. Zheng et al., 2021) to select roads based on their characteristics. Researchers also trained end-to-end DCNN and GAN approaches from raster maps directly to perform polyline generalization (Courtial et al., 2020, 2022a, 2022b; Du, Wu, Xing, et al., 2022).

GeoAI solutions for polygon-based cartographic generalization, particularly building generalization, has attracted more attention than point- and polyline-based solutions combined. Building generalization contains two steps: namely, building grouping and footprint generalization (Z. Li et al., 2004). Researchers have utilized GeoAI either to model the initial building group patterns or to perform both steps simultaneously. Random forests first were utilized for classifying buildings to determine whether they should be eliminated, retained, or aggregated (Lee et al., 2017), and later were used to classify building group patterns (He et al., 2018). DCNNs, GANs, and GCNs have been used to automate both steps of building generalization. Regarding building grouping patterns, researchers have constructed graphs to represent adjacent buildings and then employed GCNs to recognize building group patterns (Bei et al., 2019; X. Yan et al., 2020; R. Zhao et al., 2020) and classify the building group patterns (X. Yan et al., 2019).

Several researchers also have trained end-to-end DCNNs and GANs approaches for extracting and generalizing buildings from aerial and topographic maps (Feng et al., 2019, 2020; Sester et al., 2018). Others have integrated generalization knowledge and constraints for building generalization and subsequent generalization evaluation (Courtial et al., 2021; Kang et al., 2021; M. Yang, Yuan, et al., 2022).

Finally, the evaluation of map generalization remains a thorny problem (Courtial et al., 2021). Despite established first principles such as Töpfer's radical law (Töpfer & Pillewizer, 1966), there is no readily accepted global measure for generalization quality (Touya, 2012). GeoAI offers one possible pathway for improving consistency in generalization evaluation, and thus in generalization itself (Courtial et al., 2022a; M. Yang, Yuan, et al., 2022).

### *5.2. Symbolization*

Map symbolization refers to the graphic encoding of geographic information (White, 2017). In this section, we summarize GeoAI for symbolization applications into two tasks: visual variable representation and map style transfer.

The *visual variables* (e.g., color, size, shape, texture) are the fundamental graphic dimensions across which a symbol can be varied to deliver information, with some visual variables better encoding nominal, ordinal, or numerical levels of measurement (MacEachren, 1995). Choosing visual variables is an important step in map design, and thus forms a constraint that can be used by GeoAI models. Cartographers primarily utilize GeoAI approaches to manipulate the visual variables such as color (including color hue, color value, and color saturation) and shape. Regarding color, Chen et al., (2021) extracted several higher-level color palette characteristics such as order, match, harmony, discrimination, and uniformity from maps, and trained decision trees (AdaBoost) to evaluate the color quality of maps. Wu et al., (2022) utilized a multiple-constraint optimization approach and transferred color schemes from images (e.g., paintings and other maps) to maps and visualizations. Regarding shape, Yan et al., (2021) proposed a graph convolutional autoencoder model to learn the representations of building shapes. Further, Yan & Yang, (2022) developed an encoder-decoder architecture for shape encoding which can represent raster-based (DCNNs), graph-based (GCNs), and sequence-based (LSTM) features. Beyond just focusing on individual visual variables, researchers have leveraged semantic web technologies and knowledge graphs to encode cartographic visual variables. They have built a cartographic ontology of visual variables—such as shape, color, and texture—and enabled logistic reasoning of cartographic knowledge (Huang & Harrie, 2020; Mai et al., 2022b; Viry & Villanova-Oliver, 2021).

*Map style transfer* refers to the process of reproducing artistic styles from existing maps, paintings, or other visual artwork to new input map data. A *map style* contains a set of coherent and distinct cartographic design characteristics that enable the audience to obtain various visual impressions and emotional reactions (Christophe et al., 2016; Kent, 2009). Essentially, map style transfer is the task of converting a set of visual variables from existing maps to target maps, and therefore is an intersection between map symbolization and neural style transfer from computer science (A. N. Wu et al., 2022). To this end, Kang et al., (2019) successfully configured GANs to transfer map styles from existing maps and paintings to simple styled map data. Map style transfer is an increasingly popular research topic throughout the cartography community: Bogucka & Meng, (2019) transferred styles from artwork (e.g., watercolor textures) to maps; Li et al., (2021) transferred historical map styles to OSM maps; Chen et al., (2021) developed SMAPGAN to transfer Google Map styles to simple styled maps and remote sensing images; Christophe et al., (2022) transferred maps from simple-styled to complex-styled maps; Ping & Dingli, (2020) transferred styles for game maps; Jenny et al., (2021), (2022) produced shaded relief with terrain maps; and Wong et al., (2022) analyzed changes of bridges over time by transferring historical map styles to topographic maps. Several commonly used GANs architectures include Pix2Pix (Isola et al., 2017) and CycleGAN (Zhu et al., 2017). There also are several studies that adopted similar approaches for transferring remote sensing and aerial images to topographic maps, although these are outside the scope of this section and instead are discussed in Section 5.7.

Despite the success of GANs in transferring map styles, the results of the primarily raster-based output maps face inherent drawbacks. The generated map outputs may have blurred regions and ambiguous map symbols and toponyms, potentially removing visually insignificant but semantically important features while dissolving topographic relationships among geographic objects (Christophe et al., 2022; Kang et al., 2019; M. Wu et al., 2022). A possible solution is to develop vector-based style transfer approaches. For instance, Wu et al., (2022) avoided the aforementioned drawbacks of raster-based approaches by transferring color schemes between images and maps based on map features rather than pixels.

### *5.3. Typography*

*Typography* describes the placement and styling of text added to the map to clarify details otherwise removed through generalization and symbolization choices (Guidero, 2017). Typography is an integral part of establishing the subject and tone of the map , but has received relatively less exploration with GeoAI than generalization and symbolization. GeoAI has been used for automatic label placement and additional annotation, which may enhance the effectiveness of the visual hierarchy of maps. Prior studies have utilized neural networks (Pokonieczny & Borkowska, 2019) and deep learning (Y. Li et al., 2020) to determine where to best place labels around geographic objects. In addition, Harrie et al., (2022) employed DCNNs to evaluate the quality of label placement in city wayfinding maps, pointing to several future directions for GeoAI to consider label geometrics, relationships with icons, and spacing as part of automated label placement. Notably, several sophisticated rule-based systems already have been developed to support typography in digital maps based on years of cartographic research and best practices, and these systems have demonstrated high efficacy in handling various typographic-related tasks. Perhaps consequently, the integration of GeoAI in map typography has received less emphasis.

### *5.4. Map Reading*

*Map reading* is the process of discovering what is encoded in the map through human perception (Buckley & Kimerling, 2021). Map reading for GeoAI extends human perception to train computers how to decode map features. Through GeoAI-supported map reading, computers can recognize and identify feature locations, labels, and symbolized attributes from an input set of maps. Accordingly, map reading often is a necessary training step for GeoAI models used then to perform other generalization, symbolization, or typography tasks. We observed from our sampled articles two specific tasks that enable machines to read maps: object extraction and ontology construction.

*Object extraction* solutions primarily are adopted from object detection, localization, semantic segmentation, and representation learning approach in computer vision and deep learning. Researchers have utilized DCNNs to detect point-, polyline-, and polygon-based features, such as road intersections (Saeedimoghaddam & Stepinski, 2020), mountain summits (Torres et al., 2018), building footprints (Y. Chen et al., 2021b; Heitzler & Hurni, 2020; Uhl et al., 2018, 2020), road networks (Ekim et al., 2021; Jiao et al., 2022b, 2022a), bridges (C.-S. Wong et al., 2022), highway interchanges (Touya & Lokhat, 2020), shape detection (Y. Chen et al., 2021a), surface mine disturbance extents (Maxwell et al., 2020), hydrological features such as streams, rivers, wetlands, and lakes (Ståhl & Weimann, 2022; S. Wu, Heitzler, et al., 2022c, 2022b; Xia et al., 2022), and archaeological features (Garcia-Molsosa et al., 2021). Researchers also used DCNNs to classify maps with or without buildings (Uhl et al., 2020), and classify road types of historical maps (Can et al., 2021). Most studies use raster data sources such as scanned topographic maps, historical maps, and DEMs (Digital Elevation Model) as the input for object extraction (see Jiao et al., (2021) and Chiang et al., (2020) for reviews). Broadly speaking, extracting objects from raster maps is a focused case of image object detection, localization, and segmentation in cartography. Vector-based object extraction is less common, perhaps logically so given the added location and attribute information joined to point, polyline, and polygon features. One notable approach by Yang et al., (2022) uses GCNs to detect interchanges from vector road networks.

In addition to detecting map features such as points, lines, and polygons in maps, researchers also have attempted to detect labels and pictorial objects from maps. Li et al. (2018) utilized faster R-CNNs to localize map labels and query gazetteers based on map contents. Researchers also have utilized segmentation-based DCNN approaches to detect texts from historical maps (Can & Erdem Kabadayi, 2021; Z. Li et al., 2021; Weinman et al., 2019). Prior studies also have employed DCNNs to classify maps as pictorial vs. non-pictorial maps as well as detect certain objects such as sailing ships from maps (Schnürer et al., 2021).

Second, *ontology construction* describes the process of summarizing and structuring map knowledge. For manual ontology construction, cartographers define a set of concepts, categories, and relationships within related maps as a structured framework and then employ this framework to construct the ontology among maps. Once transformed into a knowledge base, maps can be interpreted coherently by both machines and humans (Varanka & Usery, 2018). Researchers have utilized knowledge graphs and semantic web technologies to automate ontology construction from maps, which then enable computational reasoning for other cartographic design decisions. Varanka & Usery (2018) illustrated the usage of GeoSPARQL for visualizing map semantics. Huang & Harrie (2020) and Viry & Villanova-Oliver (2021) employed semantic web technologies to build an ontology of map design components—including cartographic scale, styles, symbols, and legends—that allows users to retrieve cartographic and visualization knowledge. Mai et al. (2022) constructed a cartographic ontology including content, symbols, and legends, to create a narrative-based cartographic knowledge graph system. Using these approaches, machines can not only read maps, but humans also can interactively explore, analyze, and query cartographic knowledge, keeping the human-in-the-loop.

### *5.5. Map Interpretation*

*Map interpretation* refers to the explanation of relationships among map features and their corresponding features in the world through human cognition (Kimerling et al., 2016). The distinction between *map reading* and *map interpretation* is that *map reading* concentrates on certain map objects, whereas *map interpretation* is essential to extract geographic information and identify relationships among map features. Therefore, map reading can be considered as a pre-step of map interpretation. We observed two specific tasks that enable machines to interpret maps: map content inference and map feature alignment.

*Map content inference* describes identifying the geographic location and extent of a map. Evans et al., (2017) first developed a LiveMap system that could infer the locations of input maps, supported by a DCNN – ResNet approach. Subsequently, researchers utilized DCNNs to infer map scales (Touya et al., 2020), projections (J. Li & Xiao, 2019), and layer content (Hu et al., 2021; J. Li & Xiao, 2019; Tavakkol et al., 2019; Touya et al., 2020), effectively obtaining the meta-information of maps automatically using GeoAI. Further, A. C. Robinson & Zhu, (2022) used a similar approach to trace the origin source of viral maps circulated through social media.

*Map feature alignment* refers to matching map features from different maps. Aligning map features from historical maps and contemporary maps is useful for illustrating change over time. Duan et al., (2020) propose a vector-to-raster algorithm using reinforcement learning that aligns map features in historical maps with those in contemporary maps. Shbita et al., (2020) utilized knowledge graphs to construct a linked spatio-temporal data graph that matches geographic features at different times, allowing users to effectively query map features over time periods. S. Wu et al., (2022) leveraged a DCNN-based U-Net model for historical map registration that aligned historical maps at multiple time stamps.

### 5.6. Map Analysis

*Map analysis* refers to the analytical reasoning about maps following their reading and interpretation. We only observed the use of GeoAI for map analysis to classify map types, indicating a potential gap for future research merging GeoAI, cartography, and spatial analysis.

*Map type classification* refers to the identification of phenomena of geographic patterns and cartographic representations portrayed in maps. It is a prerequisite step for further map analysis and spatial analysis, as maps with different types have different analytical purposes and affordances. Researchers have utilized DCNNs to classify maps into multiple categories. Zhou et al. (2018) created a dataset and trained DCNNs to classify maps into seven types: topographic, urban, national, 3D, nighttime, orthophoto, and land maps. Yang et al. (2020) employed DCNNs to classify map themes based on the metadata, legend, and content of maps. Scheider & Huisjes (2019) utilized multiple approaches including decision trees and classified maps based on their extensive (i.e., increase with the size of their supporting object) and intensive (i.e., independent on the size of their responding map object) properties. Hosseini et al., (2022) developed an open-sourced software *MapReader* that allows users to train and test models for map type classifications based on customized labels.

Compare to map interpretation, map analysis requires a high level of cognitive ability and prior knowledge. Accordingly, sophisticated insights related to spatial relationships, dynamics, and patterns could be discoverable using GeoAI such as, for instance, identifying clustered regions and revealing flow relationships in maps (R. E. Roth, 2013). The current progress of GeoAI for map analysis has focused primarily on shallow geographic information. In the future, it is important to develop advanced GeoAI approaches to delve deeper into the analytical aspects of maps and deliver comprehensive geographic information that reveals complex spatial patterns and interactions within maps. The emergence of generative AI offers one potentially fruitful available for new research on GeoAI applications to map analysis, as discussed in Section 7.2.

### *5.7. Map Production*

Map production refers to the multiple, often iterative steps for map creation (Buckingham, 2019). Map production can be a laborious process often with rote design and editing tasks. GeoAI can facilitate different stages in the map production workflow to allow cartographers to focus on more complex or creative tasks during design. However, the primary application of GeoAI for map production in the sampled articles was for the initial stages of map data preparation.

*Map data preparation* is a prerequisite step of using GeoAI for cartography because training GeoAI requires large-scale and high-quality geospatial data. Researchers have developed approaches to either generate or enrich geospatial data for map production. Jin et al., (2021) have leveraged GeoAI to generate map tiles from remote sensing images, a process described as satellite-images–to-maps (si2map). The output of this process is a full map tileset rather than discrete map features and objects. Several novel GeoAI models have been developed for map production such as GeoGAN (Ganguli et al., 2019), SG-GAN (Y. Zhang et al., 2020), and series GANs (X. Chen et al., 2022). Researchers also achieved the conversion from maps to satellite images using similar approaches (C. Xu & Zhao, 2018; B. Zhao et al., 2021a). In addition, Yu & Chen, (2022) have leveraged an encoder-decoder structure to fill gaps of polylines which offers valuable insights for handling missing values in map data.

Prior studies also have attempted to enrich geospatial datasets using GeoAI, an application that often overlaps with map object detection. For instance, Jiao et al., (2022a) enriched a dataset by rotating road networks while preserving other features in the map such as numbers and triangulation points. By doing so, more map data for training GeoAI models can be generated. Li et al., (2021) generated text annotations on historical maps to enrich the training dataset. Hu et al., (2021) proposed a computational workflow that generates large-scale map datasets from vector data that considers multiple types of metadata (e.g., spatial extents, place names).

## 6. Social and Ethical Implications of GeoAI in Cartography

Despite the success of research integrating GeoAI for cartography, we discovered through our review that discussion of the ethical implications of GeoAI for cartography remains underdeveloped. *Ethics of GeoAI for cartography* refers to the rules of conduct for researchers, professionals, and policymakers regarding acceptable and unacceptable application of GeoAI methods for cartographic design. Here, we draw on and extend discussions of ethics from cartography, GIScience, and computer science (e.g., Crampton, 1995; Goodchild et al., 2022; Janowicz et al., 2022; Jobin et al., 2019; Nelson et al., 2022; Zhao et al., 2021). Specifically, we raise five of potentially numerous ethical challenges facing GeoAI for cartography: commodification, responsibility, privacy, bias, and (together) transparency, explainability, and provenance. We recognize that addressing every facet of ethics, especially considering the relatively nascent discussion of ethics in GeoAI for cartography, is beyond the scope of this paper. However, we want to highlight some ethical dimensions of GeoAI for cartography prompted from our review that might serve as a foundation for future research and practices on the ethics of GeoAI for cartography.

### *6.1. Commodification*

*Commodification* in cartography refers to releasing maps and map data as products that can be bought and sold on a market (Crampton, 1995). GeoAI approaches for cartography lead to at least three new map commodities: the map data and other visual artwork used for training GeoAI applications, the GeoAI models and algorithms themselves, and the cartographic output from GeoAI models (e.g., new artistic styles, evaluation metrics). Geospatial companies may monitor, collect, and even control users' behaviors with their cartographic products to enhance their profit margin (Goss, 1995). Commodification of GeoAI-enabled map products raises new ethical questions around intellectual property, ownership, and access for cartography.

On the one hand, commodification may spur technological innovation, require new safeguards for intellectual property rights, and lead to new copyright policy for GeoAI-assisted cartographic products (Kitchin & Lauriault, 2014). However, explicit guidelines and best practices on charging for access to copyrighted map products is lacking. Consider commodification under a hypothetical scenario on map style transfer. A user uploads unstyled, open source map data to a GeoAI-empowered map style transfer service along with images of paintings to use as the inspiration style. Who owns the copyright of the transferred map style? Should the service provider retain copyright for configuring the GeoAI model for public use and supplying the computing resources to employ the model? Should the user retain copyright for providing the map data and selecting the inspiration artwork; should they only retain copyright if they pay a certain fee? Finally, should the artists who created the paintings retain copyright, or at least reserve the right to license their artwork for public consumption, but without inclusion in AI training? Should the artists get royalties if their artwork is selected for style transfer? The existing copyright framework and trademark law, as well as the current open access Creative Commons (CC) license system, do not address these issues. Cartographers and GIScientists need to have a voice in this conversation to ensure new AI copyright laws cover the nuances in our profession. Another potential solution for protecting the copyright of maps may refer to the Web3 applications supported by the decentralized blockchain technology (Potts & Rennie, 2019). However, the use of blockchain in practice opens addition ethical debates and needs further examination (Mitchell et al., 2019).

On the other hand, researchers prefer to endorse open-access datasets and models that empower the replicability and reproducibility of academic studies (Kitchin & Lauriault, 2014; Wilson et al., 2021). Furthermore, it is necessary to make open-source alternatives to proprietary GeoAI models in order to mitigate a growing digital divide in their access and use. However, the majority of GeoAI models currently are used and controlled by developed countries (Xaltius, 2020), and charging for the GeoAI models and resulting cartographic products will widen the gap between developed and developing countries. Thus, it is necessary to strike a balance between premium and gratis services and products, and to advocate for equitable access to GeoAI models for cartography. Arguably, a new digital colonialism is emerging through the dominance of developed countries in AI technology, and new critical and ethical approaches are needed to rethink GeoAI for cartography as both an emerging commodity and an instrument of power (Kwet, 2019).

Finally, commodities are not made of their raw materials alone, but also comprise the labor used to make them. While GeoAI can automate some arduous tasks for cartographers so that they can focus on more creative work, is the ultimate goal of GeoAI for cartography complete automation of map design? Hopefully, the answer is no. But, as GeoAI is pursued within cartography, we need to ask what is at stake for cartography as a profession and a labor market (Dubber et al., 2020; Zarifhonarvar, 2023), and what will we lose as human beings if map design is completely overtaken by machines?

### *6.2. Responsibility*

As Griffin, p.9 (2020) asks, "How much should we trust a machine-generated map"? To ensure that machine-generated maps are trustworthy, it is essential to articulate and regulate *responsible* GeoAI for cartography, or map applications of GeoAI that are performed appropriately and minimize harmful effects on people and the environment (Jobin et al., 2019).

GeoAI for cartography has potential to do real-world social good, such as enable real-time disaster and emergency response mapping, monitor our changing climate and develop downscaled local interventions for adaptive management of climate change impacts, and reveal deeply rooted social disparities and develop more equitable solutions for redistricting, resource allocation, and governance. However, without a framework of responsible GeoAI for cartography, its application is likely to do as much or more harm than good (Elwood & Wilson, 2017; Janowicz et al., 2022; Sieber, 2006). To limit abuse of GeoAI in cartography, it is necessary to recognize who controls the map data, the cartographic narrative, and the interpretation of maps. GeoAI models, just like maps, reflect the interests and viewpoints of their makers, and the application of GeoAI for cartography thereby escalates existing power dynamics. For instance, what could Participatory GeoAI look like, taking inspiration from public participation GIS (PPGIS), participatory mapping, counter mapping, and collaborative mapping (Bosse, 2021; Chambers, 2006; Fagerholm et al., 2021; Peluso, 1995). Such efforts may both clarify responsibility for and democratize control over GeoAI for cartography, ensuring that maps that are created and interpreted by GeoAI do not solely reflect the perspectives of those in power, but also involve (without abusing) the voices of marginalized and underrepresented communities.

Also, who should be responsible for the AI’s metaphorical “choices” when producing maps? In particular, who should be responsible if GeoAI models produce “unethical” maps that are inaccurate, misleading, propagandized, or even used for illegal activities? Answering these questions is challenging because formally defining an ethical versus unethical map is challenging, and the definition may take different shapes to researchers, professionals, and policymakers and in different mapping contexts. For instance, the *GIS Certification Institute Code of Ethics* (GIS Certification Institute, 2022) outlines four “obligations” for professionals that read much like responsibilities: obligations to society, obligations to employers and funders, obligations to colleagues and the profession, and obligations to individuals in society (relating to geoprivacy below in Section 6.3). What is missing from these obligations when considered GeoAI? Researchers may be quick to point out these obligations do not require any deep engagement with the places and communities that are being mapped, an issue of extractive scholarship well-documented in cartography that GeoAI only exacerbates through scaling up the speed and coverage of such extractive practices. Similarly, policymakers may be quick to point out that these obligations focus on people rather than the environment and its non-human inhabitants, and also do not require professionals to take actions beyond delivering the map. GeoAI may even complicate evidence-based policy and informed action given the potential for “fake maps” (B. Zhao et al., 2021a).

Further, any delineation of an ethical versus unethical map will shift through time and across communities and cultures. For instance, Holloway's (2007) *Right Mapmaking: Five Ways to Make Maps for a Future to Be Possible* provides an alternative to the aforementioned professional obligations, centering on five “percepts” that drawn on multicultural ethos: reverence, the practice of generosity, commitment to the relationship with place, deep listening through direct-contact and stopping, and on belong to one body. These percepts are loosely connected to Hindu, Christian, Buddhist, Taosit, and Muslim morals, respectively, and therefore provide a potential model example of multicultural hybridity.

### *6.3. Geoprivacy*

*Geoprivacy* refers to the protection of individuals' right to protect their locational information from unwanted disclosure (Kounadi & Leitner, 2014; Kwan et al., 2004; Weidemann et al., 2018). As discussed above, GeoAI models are trained on map data from authoritative, commercial, and user-generated sources, each of which could contain confidential locational information. Privacy is an acute concern for user-generated data sources, as individuals may be contributing large volumes of spatial data unconsciously through social media and other location based services (Harvey, 2013; Thatcher et al., 2018). In an era of "big data", individuals may feel anonymous when volunteering information given the sheer volume and velocity of information posted to these streams. However, GeoAI models can make sense of big data streams larger than any single human can closely review, and thus might disclose intimate trajectories, activities, and behaviors that are previously concealed in the data, undermining individual geoprivacy. How do we educate users about new privacy concerns derived from GeoAI before they volunteer their data, and how do we prevent GeoAI from revealing private locational information when used for cartography?

While user-generated map data may be the most obvious concern regarding geoprivacy, GeoAI also has potential to reveal private information in both authoritative and commercial data sources. Specifically, authoritative and commercial map data typically are aggregated or anonymized to preserve geoprivacy before publishing the dataset (e.g., Kounadi & Leitner, 2014). However, government and industry are not prevented from training their GeoAI models on individual-level data internally before publishing it for third-party usage. Arguably, the potential for nefarious misuse of private information by GeoAI actually is greater for authoritative and commercial map data, in part because it is more difficult to monitor and research from outside these institutions. Perhaps the most high-profile example with authoritative data is predictive policing, an application of GeoAI for cartography that has received fair criticism as a new form of racial and geographic profiling (Jefferson, 2018; Lally, 2022). With commercial data, private locational information may become a service that can be purchased by marketers, making the users' lives a product for sale (a fourth commodity, tying into the above discussion about commodification in Section 6.1). Geoprivacy requires that we ask "Who truly owns the big data?" and therefore who has the right to utilize these big data streams in GeoAI applications for cartography. For instance, what should happen to individual locational data when a company goes bankrupt (leaving no funding to manage sensitive information) or merges with a parent company (with potentially different operating policies and standards to which users previously agreed). When can authoritative, commercial, or user-generated map data be subpoenaed in a trial or requested as a term of employment? Answers to these questions currently vary considerably by geography. Finally, how can GeoAI models be more transparent about how potentially-sensitive information is used while training the model (see additional discussion below in Section 6.5).

### *6.4.Bias*

*Bias* describes the systematic errors, inaccuracies and misinformation, and skewed representations that are possibly present in the data, models, algorithms, or interpretations of GeoAI for cartography. All maps represent data that are subject to uncertainty (Couclelis, 2003), and these map data may vary in their accuracy, precision, currency, completeness, consistency, etc. (MacEachren et al., 2012). GeoAI models amplify these biases when learning from uncertain data during training and then applying biased patterns to output cartographic products. GeoAI approaches also may introduce new data biases when creating cartographic outcomessuch as representation bias, measurement bias, and evaluation bias (Jobin et al., 2019; Miller, 1995; Nelson et al., 2022)— nd exhibit biases within their algorithmic structures, (Kang, Abraham, et al., 2023; Roselli et al., 2019; J. Wu et al., 2023; Zou & Schiebinger, 2018) since GeoAI models are developed by humans who themselves hold biases.

Bias also exists in where and who the training dataset includes. For instance, the quality of a crowdsourced dataset may vary across different regions. Specifically, the dataset might be sparse in rural areas compared to urban areas, which may cause an imbalance in the data representation (G. Zhang & Zhu, 2018). The dataset also may privilege particular groups of people by gender identity, race and ethnicity, sexual orientation, physical and cognitive ability, age, education level, socioeconomic status, and cultural background, among others (Stephens, 2013), leading to generated map products that work better for those holding power (D'Ignazio & Klein, 2016; Kelly, 2021). In constrast, the dataset might target marginalized groups of people due to historical or institutional reasons, leading the GeoAI model to overfit (and thus incorrectly predict) particular spatio-demographic patterns, such as the aforementioned case of predictive policing (Jefferson, 2018; Lally, 2022).

Therefore, it is necessary to observe, analyze, and mitigate bias in GeoAI applications for cartography. One potential solution is to train a GeoAI model that could be used to help humans assess different kinds of observed biases in modeled outcomes. Further, a potentially fruitful area of future research is to bring human-in-the-loop GeoAI to cartography to order to help observe and mitigate different forms of biases. Ultimately, researchers and practitioners in GeoAI for cartography should formalize their representational practices at the onset of a new project (e.g., Brown & Knopp, 2008; Kelly, 2021; Kirby et al., 2021; Pavlovskaya & Martin, 2007). They should prioritize marginalized use cases at all decision points in a project. It emphasizes not only the input training data and output map products but also who should and should not make use of the results and for what purpose. D'Ignazio & Klein, (2020) offer seven data feminism principles that can serve as a guiding framework for addressing bias in GeoAI for cartography: examine power; challenge power; elevate emotion and embodiment; rethink binaries and hierarchies; embrace pluralism; consider context; and make labor visible. Arguably, each of these data feminism principles is a social and ethical implication in GeoAI for cartography beyond simply bias. Notably, early work in data feminism emphasizes application of these ethical guides to both design process and output (D'Ignazio & Klein, 2016), making transparency, explainability, and provenance (treated below) of the *process* an ethical imperative for GeoAI as applied to cartography and beyond.

### *6.5. Transparency, Explainability, and Provenance*

*Transparency* and *explainability* together describe the ability to decompose, understand, and communicate the behaviors of a GeoAI model. Given the application of GeoAI to cartography, this also includes the ability to visualize or otherwise cartographically represent the map output at different stages of the modeling process.

Current AI models, especially deep learning approaches, often are described as "black-boxes" given that the researchers, professionals, or policymakers who use them may not, or, in many cases, cannot uncover the mechanism behind the output of models (Ricker, 2017). In these AI models, different parameters and initialization states produce widely different outcomes, making it difficult to gauge sensitivity of the model to these settings as well as to reproduce consistent results. Transparency and explainability is essential for establishing responsibility and bias in the GeoAI models and building public trust in the cartographic output.

Making AI models transparent and explainable leads to the issue of *provenance*, or the ability to track GeoAI "choices" while the model is running. Provenance is particularly important for understanding how GeoAI models produce maps deemed as unethical, both to improve the underlying model for future use and to identify intentional misuses of the model. However, recording and retracing AI behaviors remains a major challenge. Relatedly, GeoAI techniques are evolving at a breakneck pace: how do we archive previous versions of a model or algorithm and track updates over time? At what point should a cartographic output be deprecated or "re-run" given updates to the underlying GeoAI model? Again, GeoAI could be used to track, document, and interpret the provenance of GeoAI models and outcomes for cartography.

# 7. Discussion

## *7.1. Takeaways of GeoAI for Cartographic Design Decisions*

Beyond specific models, applications, and ethical questions, we observed several overarching patterns from the review and synthesis. First, we noticed that cartographic design decisions that are primarily related to the map's "mechanics" or basic construction, such as map generalization, map reading, and map production, have garnered the most interest in the research literature to-date and appear to be the closest to professional automation. In contrast, more artistic and creative cartographic design decisions, such as symbolization and typography, have received relatively less attention, and the limited studies on these design topics have a relatively narrow focus (e.g., visual variable representation, map style transfer). GeoAI is particularly adept at solving "engineering" problems, and therefore may hold more immediate benefit for map construction over more creative aspects of the cartographic design process.

Further, there are a range of cartographic design decisions that have so far yielded minimal interest from the GeoAI research community. Among entries listed in the *Map Design Fundamentals* section of the *GIS&T Body of Knowledge (2022)*, there were no papers in our sample on *Statistical Mapping*, *Map Projections*, *Visual Hierarchy and Layout*, *Color Theory*, or *Design and Aesthetics*. Accordingly, one potentially fruitful future research direction is employment of GeoAI on these cartographic design decisions using representation learning (Bengio et al., 2013). For example, cartographic design characteristics like projections, color schemes, visual hierarchy, and layout can be encoded as high-dimensional visual vectors that contain essential artistic and geographic information of maps, moving past map mechanics to explore the more artistic dimensions of map design. Another possible research direction is the development of a cartographic design recommendation system based on GeoAI. For instance, recommendations could include optimal classification approaches for choropleth maps, appropriate projections for different thematic map types at different scales, improved layouts to make better use of negative space, color palettes that work for different kinds of color vision deficiency, and so on. Such a GeoAI-enabled recommendation system built on cartographic expertise would help untrained mapmakers create reasonably-designed maps, making it easier for anyone to make their own maps and improving the communication quality of maps overall.

### *7.2. Future Directions of GeoAI for Cartography*

In the addition to the above takeaways, we identified four key gaps between the existing literature on GeoAI for cartography and trends in AI broadly: GeoAI-enabled active cartographic symbolism, human-in-the-loop GeoAI for cartography, GeoAI-based mapping-as-a-service, and generative AI for cartography.

First, GeoAI can enable *active symbolism*, or the use of GeoAI by cartographers to generate and evaluate alternative cartographic design options that meet specific map use and user contexts (Armstrong, 2019). Several early attempts to automate cartographic design through intelligent systems took a *rule-based* approach, which distills cartographic expertise and time-tested recommendations into a set of pre-defined rules for use in a decision tree, optimization function, etc. (e.g., Buttenfield & Freeman, 1991; Jones & Mark Ware, 2005; Kang, 2020; Yan et al., 2019). While cartography has developed a canon of principles to inform design, the actual cartographic design process is far more non-linear and iterative than a rule-based system suggests (Nestel, 2019). Emerging

GeoAI models do not need input rules, however, and instead take a *pattern-based* approach to learn cartographic principles and subsequently apply these patterns to new maps (Kang, 2020; X. Yan et al., 2019). Thus, GeoAI can serve an exploratory role for cartographers, using different inspiration input maps and artwork to break from dominant conventions in cartographic design. Arguably, map style transfer has gotten the closest to active cartographic symbolism to-date (Christophe et al., 2022; Kang et al., 2019); however, this approach could support creative brainstorming across the cartographic design decisions discussed in the *Cartography and Visualization* section of the *GIS&T Body of Knowledge* (2022). Active symbolism also can be used to evaluate differences in cartographic design across regions and cultures, and even adapt design of digital maps as they move across place on mobile devices to improve effectiveness and sensitively for local contexts (R. Roth et al., 2023).

Second, human-in-the-loop GeoAI has the potential to further empower cartographers (Zanzotto, 2019), going a step beyond active symbolism that only enables humans at the output stage to make the designer a participant in decision making as the model is running. Human-in-the-loop GeoAI relates to *model steering*, an early initiative of visualization in scientific computing (McCormick, 1988), and now core tenet of geovisual analytics, with the goal of supplying map interfaces to computational methods in support of sophisticated analytical reasoning (Andrienko et al., 2007). Human-in-the-loop GeoAI keeps decision-making authority in the hands of the cartographer, allowing them to bend the output design to their needs and preferences. Human-in-the-loop GeoAI for cartography also represents a major step towards transparency, explainability, and provenance, and therefore has important ethical implications as well. Finally, putting the human-in-the-loop has the potential to save computational resources, as meaningless or suboptimal designs can be abandoned early in the model processing, aligning with green cartography initiatives to reduce energy consumption and the associated carbon footprint (Han et al., 2021).

Third, GeoAI can support mapping-as-a-service, shifting focus from the map as a single product and to the map as a platform supporting online map design (Overstall, 2021). Proprietary mapping-as-a-service platforms such as ArcGIS Online, Carto, Felt, and Mapbox Studio have replaced traditional desktop GIS for many mapping needs, enabling users to collate map data, customize basemap tilesets, and produce and share simple thematic maps. GeoAI combined with cloud services have the potential to enhance mapping-as-a-service platforms with advanced spatial analytics as well as supply the aforementioned design recommendations for beginning users. Notably, mapping-as-a-service exacerbates the ethical challenge of commodification, as the costs associated with serving GeoAI-based map services online may prohibit competitive open source solutions, leading to "freemium" models that restrict access to advanced features without a subscription.

Finally, the emergence of generative AI presents both new opportunities and new ethical challenges for cartography. *Generative AI* models such as ChatGPT[2] and DALL·E 2[3] have received significant public attention due to their ability to communicate in natural language and generate realistic images and, potentially, realistic-looking maps (Kang, Zhang, et al., 2023). These generative AI models usually are supported by several large language models (LLM), and enable text-to-text (e.g., ChatGPT) or text-to-image generation applications (e.g., DALL·E 2). Some models even allow image-to-text generations (J. Li et al., 2023; Schuhmann et al., 2022). Other generative AI include Google's Bard[4], Midjourney[5], and Stable Diffusion[6]. Notably, during the timeframe of our review and synthesis (before December 2022), we did not observe mention or application of any of these generative AI models for cartography given the recency of their development. Since, several exploratory projects have emerged and have demonstrated the current possibilities and limitations of generative AI for cartographic applications (Kang, Zhang, et al., 2023; Tao & Xu, 2023). While it is impossible to predict where generative AI will be in 5, 10, or 25 years, it is very possible that our review and synthesis presented outlines the history of GeoAI for cartography *before* a new era of generative GeoAI. While these generative AI models hold the potential to fundamentally change cartography and GIScience, they also pose new ethical concerns around commodification, responsibility, geoprivacy, bias, and transparency/explainability/provenance. At the time of this writing, generative AI has sparked debates regarding its social and ethical implications such as future employment prospects (Zarifhonarvar, 2023), trustworthiness (Tlili et al., 2023), and potential negative effects of unrestricted technological development (King & chatGPT, 2023; van Dis et al., 2023). Therefore, it is necessary to carefully consider the potential ethical issues related to the generative AI in cartography. For instance, Kang et al., (2023) have discussed several potential ethical concerns that may arise from AI-generated maps including inaccuracies, misleading information, unanticipated features, and irreproducibility.

[2] https://openai.com/blog/chatgpt

[3] https://openai.com/dall-e-2

[4] https://bard.google.com/

[5] https://www.midjourney.com/

[6] https://stablediffusionweb.com/

### *7.3. Related Research Foci*

We framed our synthesis and review as "GeoAI for cartography", but, as a nascent research thrust, even this phrasing may evolve over time. Since we completed the review in December 2022, similar research has been framed as MapAI (A. C. Robinson, 2023) and CartoAI (Feng et al., 2023). Is there a set of objectives and techniques common among GeoAI for cartography, CartoAI, and MapAI, or do these possibly represent different research perspectives? Further, ambiguity remains between the terms AI and GeoAI: if all or most data are spatial, does that mean all AI is GeoAI? In this section, we engage with these concepts to stimulate further discussion and reflection about the concordances and dissonances among these terms.

Janowicz et al., (2020) introduced Geospatial Artificial Intelligence (GeoAI) as "*a subfield of spatial data science utilizes advancements in techniques and data cultures to support the creation of more intelligent geographic information as well as methods, systems, and services for a variety of downstream tasks*", and Gao, (2021) defined GeoAI "*to develop intelligent computer programs to mimic the processes of human perception, spatial reasoning, and discovery about geographical phenomena and dynamics; to advance our knowledge; and to solve problems in human environmental systems and their interactions, with a focus on spatial contexts and roots in geography or geographic information science (GIScience)*". Both definitions are intertwined with the concept of artificial intelligence (AI) but with a special focus on geographic components. Broadly speaking, any applications of AI that address geographic problems belong to GeoAI, although the perhaps the ultimate promise of GeoAI is development of spatially explicit models, i.e., empowering AI with spatial thinking and reasoning capabilities.

MapAI and CartoAI originate from the names of two conference workshops that specifically focused on the utilization of AI for cartography. MapAI explores the "*ideation and practical experimentation collaboratively explore some of the potential and limits of current AI technologies for cartographic practice and map use*" (A. C. Robinson, 2023), and CartoAI is defined as "*artificial intelligence techniques applied to cartography*" (Feng et al., 2023). These definitions underscore the applications of AI in cartography. Beyond their original conceptions, we propose that these two terms may contain more expansive implications for future studies. Specifically, MapAI could potentially include broader contexts such as the integration of AI in high-resolution maps utilized in autonomous vehicles, the utilization of AI in map making from remote sensing images. These applications may not inherently involve cartographic design principles. In comparison, the concept of CartoAI might focus more on the encoding of cartographic principles within AI models, thereby addressing complex cartographic design decisions. Such a focus includes more aesthetic and artistic dimensions of cartography, reflecting a synthesis of design, functionality, and innovation in map creation and representation.

In sum, this paper concentrates on the applications of GeoAI for cartography, and highlights how GeoAI can be leveraged to support cartographic design decisions. Terms including GeoAI for cartography, AI for cartography, CartoAI, and MapAI might be interchangeable for most contexts, especially as their differences are quite subtle at the moment. However, as the cartography studies and AI technologies continue to evolve, precise definitions and differentiations might be needed in future studies to distinguish them.

### *7.4.Limitations*

Despite the success and promise of integrating GeoAI approaches to support cartographic design, they are still far from perfect. Similarly, the review and synthesis reported here has several limitations. First, despite having produced promising results, a large proportion of existing GeoAI technologies are still far from being applied to real-world cartographic applications. For example, certain GeoAI models may struggle to preserve spatial and topological structures in maps, leading to the loss of important geographic objects, blurred regions, and distorted cartographic outputs (Courtial et al., 2020; Griffin, 2020; M. Wu et al., 2022). Thus, just because we summarize GeoAI for cartography in research contexts does not mean these processes are transferable to professional design yet.

Second, even though it is necessary to involve the human-in-the-loop throughout the development and evaluation of GeoAI, different people may have distinct preferences and tastes regarding map design. Moreover, best practices and conventions in cartography evolve over time, and we acknowledge there are many reasons to break from these best practices and conventions with any given map. Existing GeoAI approaches often rely on learning average patterns from large-scale datasets and neglect unique or novel styles that actually might reflect elite design in cartography. Thus, GeoAI may never be able to fully realize the diverse map design derived from professional cartographers, or innovate design beyond the status quo.

Finally, while we attempted to provide a comprehensive overview of GeoAI in cartography, it is necessary to acknowledge that there might be instances of overlooked papers. We only included papers written in English and similarly present this review in English only given our own constraints as scholars working at English-speaking universities. However, additional research on GeoAI for cartography has been published in Chinese (Ai, 2021; Wang et al., 2022), Russian (Poshivaylo & Kolesnikov, 2021), and other languages. Some literature, including conference papers not indexed by Scopus, might be absent from our review, although the backwards search did help us identify relevant papers in the *International Cartography Conference* (ICC) and *AutoCarto* proceedings. Further, despite that such keyword-based search strategy has been widely adopted and utilized in literature review studies, it is necessary to acknowledge several broader issues associated with reliance on algorithms for literature selection that potentially narrow the scope of the reviewed papers. Hence, we advocate for more critical and reflexive approaches in future review papers as GeoAI for cartography grows.

## 8. Conclusion

In this article, we presented a comprehensive content analysis and narrative synthesis of research studies that integrated GeoAI in cartography. We outlined current research and advancements in the usage of GeoAI for cartographic design. To do so, we first introduced data sources, data formats, GeoAI methods, and evaluations of integrating GeoAI in cartography. Notably, six primary GeoAI models are demonstrated including decision trees, knowledge graph and semantic web technologies, deep convolutional neural networks, generative adversarial networks, graph neural networks, and reinforcement learning. We then summarized seven applications of GeoAI for cartographic design decisions including generalization, symbolization, typography, map reading, map interpretation, map analysis, and map production. After that, we raised five key ethical challenges regarding GeoAI for cartography: commodification, responsibility, privacy, bias, and (together) transparency, explainability, and provenance. Finally, we proposed and discussed four promising research directions in this emerging field including GeoAI-enabled active cartographic symbolism, human-in-the-loop GeoAI for cartography, GeoAI-based mapping-as-a-service, and generative GeoAI for cartography.

Our study offers valuable insights and contributions for researchers, professionals, and policymakers. For researchers, we demonstrate the current progress of GeoAI for cartography and offer a comprehensive illustration of how GeoAI could support different cartographic design decisions. For professionals, our review guides best practices regarding how GeoAI models and methods can be leveraged during cartographic design to facilitate their mapmaking processes. For policymakers and researchers in critical cartography and GIS, we emphasize the significance of integrating responsible and ethical practices when drawing on GeoAI for cartographic design. By highlighting several ethical implications, we contribute to discussions regarding the development of equitable and inclusive GeoAI technology.

At the end of this review, we raise a pivotal question that worth further discussion in the future: "When should we *not* use GeoAI for cartography?" Reflecting upon this question may not only facilitate the development of ethical GeoAI for cartography, but also provide a better understanding of the practical scope and capabilities of GeoAI in cartography. We invite our readers to participate in this discussion, as the integration of a wide range of perspectives is vital to assure the ethical development and use of GeoAI in cartography.

## Data Availability Statement

The authors confirm that the data supporting the findings of this study are available within the article and its supplementary materials.

## Acknowledgments

The authors would like to thank people who attended the AutoCarto 2022 Conference and CartoAI workshop at the GIScience 2023 conference. The feedback and insights from those who participated the discussions have been valuable for this paper. In particular, Dr. Liqiu Meng, Mark Cygan, Dr. Alexander Kent, and Dr. Feng Yu. The authors would also like to acknowledge the anonymous reviewers whose critical reviews and suggestions have helped improve the quality of this paper. This work is supported by the Trewartha Research Award at the University of Wisconsin-Madison, Master's Thesis Research Grant of the AAG Cartography Specialty Group. During the writing of the manuscript, ChatGPT was utilized as a tool solely for proofreading purposes, without contributing any ideas or perspectives to the content of the paper.